\documentstyle[amssymb,prl,aps,twocolumn,epsf]{revtex}

\newcommand{\be}{\begin{equation}}
\newcommand{\ee}{\end{equation}}
\newcommand{\bea}{\begin{eqnarray}}
\newcommand{\eea}{\end{eqnarray}}

\begin{document}

\twocolumn[\hsize\textwidth\columnwidth\hsize\csname @twocolumnfalse\endcsname

\title {Fingerprints of spin-fermion pairing in cuprates}
\author{Ar. Abanov$^1$, Andrey V. Chubukov$^1$, and  
J\"org Schmalian$^2$}   
\address{   
$^1$ Department of Physics, University of Wisconsin, Madison, WI 53706}     
\address{$^2$ Department of Physics and Ames Laboratory, 
 Iowa State University, Ames, IA 50011}   
\date{\today}
\draft
\maketitle
\begin{abstract}
We demonstrate that  the feedback effect
from bosonic excitations on fermions, 
which in the past allowed one to verify the phononic 
mechanism of a conventional, $s-$wave superconductivity,  may
also allow one to experimentally detect the ``fingerprints'' of 
the pairing mechanism in cuprates.
We argue that for spin-mediated $d-$wave superconductivity, 
the fermionic spectral function, 
the density of states, the tunneling conductance through an insulating 
junction,  and the optical conductivity are affected 
by the interaction with collective spin excitations, which below $T_c$
 are propagating, magnon-like quasiparticles
with  gap $\Delta_s$.
 We show  that the interaction with a propagating spin excitation 
gives rise to singularities at frequencies $\Delta + \Delta_s$ 
for the spectral function and the density of states, and at 
 $2\Delta + \Delta_s$ for  tunneling and optical conductivities, where
$\Delta$ is the maximum value of the $d-$wave gap. 
We further argue that  recent
optical measurements also allow one to detect
 subleading singularities at $4\Delta$ and $2\Delta + 2\Delta_s$. 
We consider  the experimental detection of  these singularities
 as a strong evidence in favor of the 
magnetic scenario for superconductivity in cuprates.
\end{abstract}
\pacs{PACS numbers:71.10.Ca,74.20.Fg,74.25.-q}
] \narrowtext

\section{Introduction}

One of the very few accepted facts for  high-$T_{c}$ materials is that they
are d-wave superconductors\cite{WVH93,TK94,electrond}. This salient
universal property of all cuprates entails strong constraints on the
microscopic mechanism of superconductivity. However, it does not uniquely
determine it, leading to a quest for experiments which can identify
''fingerprints'' of a specific microscopic theory of $d$-wave
superconductivity, a strategy somewhat similar to the one used in
conventional superconductors (see e.g., \cite{mahan}). There, the
identification of characteristic phonon frequencies in the tunneling density
of states (DOS) below $T_{c}$ 
was considered as a decisive evidence for the electron-phonon mechanism for
superconductivity\cite{Scalapino69}.

In this paper, we assume {\em a'priori} that the pairing in cuprates is
mediated by the exchange of collective spin excitations. It has been
demonstrated both numerically and analytically that this exchange gives rise
to a $d$-wave superconductivity~\cite{pairing}. We discuss to which extent
the ``fingerprints'' of the spin-mediated pairing can be extracted from the
experiments on hole-doped 
high $T_{c}$ materials. We argue that due to strong
spin-fermion coupling, there is a very strong feedback from spin excitations
on fermions, specific to $d-$wave superconductors with magnetic pairing
interaction. The origin of this feedback is the emergence of a propagating
collective spin bosonic mode below $T_{c}$. We show that this mode is
present for any coupling strength, and its gap $\Delta _{s}$ is smaller than
the minimum energy $\thicksim 2\Delta $ which is necessary to break a Cooper
pair. In the vicinity to the antiferromagnetic phase, $\Delta _{s}\propto
\xi ^{-1}$ where $\xi $ is the magnetic correlation length. We show that the
spin propagating mode changes the onset frequency for single particle
scattering, 
and gives rise to the ``peak-dip-hump'' features in angular resolved
photoemission (ARPES) experiments, the ``dip-peak'' features in tunneling
experiments, and to the singularities and fine structures in the optical
conductivity. We demonstrate that (i) these features have been observed~\cite
{Norman97,shennat,Fedorov99,Kaminski00,fisher,zasad,basov,CSB99} (ii) ARPES~ 
\cite{Norman97,shennat,Fedorov99,Kaminski00}, tunneling~\cite{fisher,zasad},
and conductivity data~\cite{basov,CSB99} are consistent with each other, and
(iii) the value of $\Delta _{s}$ extracted from these various experiments
agrees well with the resonance frequency measured directly in neutron
scattering experiments~\cite{neutrons,dai,neutrons2}.

\subsection{The physical origin of the effect}

The physical effect which accounts for dips and humps in the density of
states and spectral function of cuprates by itself is not new and is known
for conventional $s-$wave superconductors as a Holstein effect \cite
{Scalapino69,holstein,varma} 
\begin{figure}[tbp]
\begin{center}
\epsfxsize=3.2in \epsfysize=0.6in
\epsffile{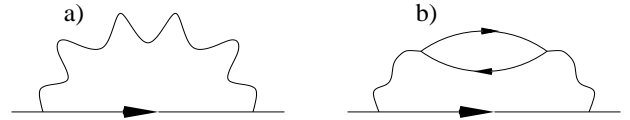}
\end{center}
\caption{a) The exchange diagram for boson mediated interaction. The solid
line stands for a propagating fermion. The wiggled line is a phonon
propagator in case of electron- phonon interaction, and a magnon line in
case of spin- fluctuation mediated interaction. b) The lowest order diagram
for the fermionic self energy due to a direct four fermion interaction, also
represented by a wiggly line}
\label{Figure1}
\end{figure}

Consider a clean $s-$wave superconductor, and suppose that the residual
interaction between fermions occurs via the exchange of an Einstein phonon.
Assume for simplicity that the fully renormalized electron phonon coupling
is some constant $g_{{\rm ep}}$, and that the phonon propagator $D(q,\omega )
$ is independent on the momentum $q$ and has a single pole at a phonon
frequency, $\Delta _{p}$ (a Holstein model) 
~\cite{holstein,varma,ssw}. The phonon exchange gives rise to a fermionic
self-energy (see Fig~\ref{Figure1}a) 
\begin{equation}
\Sigma (\omega _{m})=i\omega _{m}+g_{{\rm ep}}^{2}T\sum_{n}\int d^{3}kG_{%
{\bf k}}(\omega _{n})D(\omega _{m}-\omega _{n})  \label{s}
\end{equation}
which is a convolution of $D(\omega )=1/(\Delta _{p}^{2}-\left( \omega
+i\delta \right) ^{2})$ with the full fermionic propagator $G_{k}(\omega )$.
(For further convenience, we absorbed a bare $i\omega _{m}$ term into the
definition of $\Sigma (\omega _{m})$). In a superconductor, the fermionic
propagator is given by 
\begin{equation}
G_{{\bf k}}(\omega )=\frac{\Sigma (\omega )+\varepsilon _{{\bf k}}}{\Sigma
^{2}(\omega )-\Phi ^{2}(\omega )-\varepsilon _{{\bf k}}^{2}}
\end{equation}
where $\Phi (\omega )$ is the anomalous vertex function, and $\varepsilon _{%
{\bf k}}$ is the band dispersion of the fermions. The superconducting gap,
introduced in the BCS theory, is the solution of $\Sigma (\Delta )=\Phi
(\Delta )$.  [Alternatively to $\Sigma (\omega )$ and $\Phi (\omega )$, one
can introduce the complex effective mass function, $Z(\omega )=\Sigma
(\omega )/\omega $, and the complex effective gap function $\Delta (\omega
)=\Phi (\omega )/Z(\omega )$~ \cite{ssw,Eliashb}]. In what follows we will
use $\Sigma ^{\prime }(\omega )$, $\Phi ^{\prime }(\omega )$, etc. to denote
real parts and $\Sigma ^{\prime \prime }(\omega )$, $\Phi ^{\prime \prime
}(\omega )$, etc. for the imaginary parts of the functions we study.

For $T=0$ one can rigorously prove that both $\Sigma ^{\prime \prime
}(\omega )$ and $\Phi ^{\prime \prime }(\omega )$ vanish for $\omega \leq
\Delta $. This implies that the fermionic spectral function $A_{{\bf k}%
}(\omega )=\left| G_{{\bf k}}^{\prime \prime }(\omega )\right| /\pi $ for
particles at the Fermi surface (${\bf k}={\bf k}_{{\rm F}}$) has a $\delta-$%
functional peak at $\omega =\Delta $, i.e. $\Delta $ is a sharp gap in the
excitation spectrum at $\ $zero temperature. Also, the fermionic density of
states in a superconductor 
\begin{equation}
N(\omega )={\rm Im}\left[ \frac{\Sigma (\omega )}{(\Phi ^{2}(\omega )-\Sigma
^{2}(\omega ))^{1/2}}\right]  \label{dos}
\end{equation}
vanishes for $\omega <\Delta $ and has a square-root singularity $N(\omega
)\propto (\omega -\Delta )^{-1/2}$ for frequencies above the gap, $\omega
\geq \Delta $.

The onset of the imaginary part of the self-energy due to inelastic single
particle scattering can be easily obtained by applying \ the spectral
representation to Eq. (\ref{s}) and re-expressing the momentum integration
in terms of an integration over $\varepsilon _{{\bf k}}$. At $T=0$ we then
obtain 
\begin{equation}
\Sigma ^{\prime \prime }(\omega >0)\propto \int_{0}^{\omega }d\omega
^{\prime }N(\omega ^{\prime })D^{\prime \prime }(\omega -\omega ^{\prime })
\label{im}
\end{equation}
Since for positive frequencies, $D^{\prime \prime }(\omega )=(\pi
D_{0}/2\Delta _{p})\delta (\omega -\Delta _{p})$, the frequency integration
is elementary and yields 
\begin{equation}
\Sigma ^{\prime \prime }(\omega >0)\propto N(\omega -\Delta _{p}).  \label{1}
\end{equation}
We see that the single particle scattering rate is directly proportional to
the density of states shifted by the phonon frequency. Clearly, the
imaginary part of the fermionic self-energy emerges only when $\omega $
exceeds a threshold at 
\begin{equation}
\omega _{0}\equiv \Delta +\Delta _{p},
\end{equation}
i.e., the sum of the superconducting gap and the phonon frequency. Right
above this threshold, $\Sigma ^{\prime \prime }(\omega )\propto (\omega
-\omega _{0})^{-1/2}$. By Kramers-Kronig relation, this nonanalyticity
causes an analogous square root divergence of $\Sigma ^{\prime }(\omega )$
at $\omega <\omega _{0}$. Combining the two results, we find that near the
threshold, $\Sigma (\omega )=A+C/\sqrt{\omega _{0}-\omega }$ where $A$ and $%
C $ are real numbers. By the same reasons, the anomalous vertex $\Phi
(\omega ) $ also possesses a square-root singularity at $\omega _{0}$. Near $%
\omega =\omega _{0}$, $\Phi (\omega )=B+C/\sqrt{\omega _{0}-\omega }$ with
real $B$. Since $\omega _{0}>\Delta $, we have $A>B$.

\begin{figure}[tbp]
\begin{center}
\epsfxsize=3.0in \epsfysize=1.6in
\epsffile{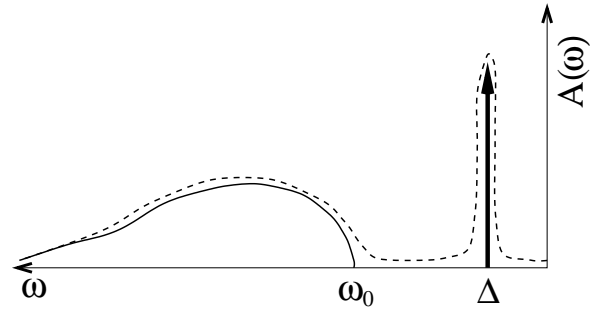}
\end{center}
\caption{The schematic form of the quasiparticle spectral function in an $s-$%
wave superconductor. Solid line -- $T=0$, dashed line -- a finite $T$. $%
\protect\omega _{0}=\Delta +\Delta _{p}$}
\label{Figure2}
\end{figure}

The singularity in the fermionic self-energy gives rise to an extra dip-hump
structure of the fermionic spectral function at ${\bf k}={\bf k}_{F}$. Below 
$\omega _{0}$, the spectral function is zero except for $\omega =\Delta $,
where it has a $\delta -$functional peak. Above $\omega _{0}$, $\ A(\omega
)\propto {\rm Im}(\Sigma (\omega )/(\Sigma ^{2}(\omega )-\Phi ^{2}(\omega ))$
emerges as $A(\omega )\propto (\omega -\omega _{0})^{1/2}$. At larger
frequencies, $A(\omega )$ passes through a maximum, and eventually vanishes.
Adding a small damping due to either impurities or finite temperatures, one
obtains the spectral function with a peak at $\omega =\Delta $, a dip at $%
\omega \approx \omega _{0}$, and a hump at a somewhat larger frequency. This
behavior is schematically shown in Fig.~\ref{Figure2}. 

The singularities in $\Sigma (\omega )$ and $\Phi (\omega )$ affect other
observables such as fermionic DOS, optical conductivity, Raman response, and
the SIS tunneling dynamical conductance~\cite{varma,coffey}.

For a more complex phonon propagator, which depends on both frequency and
momentum, actual singularities in the fermionic self-energy and other
observables are weaker and may only show up in the derivatives over
frequency~\cite{McMillan69}. 
Still, however, the opening of the new relaxational channel at $\omega _{0}$
gives rise to singularities in the electronic properties of an $s-$wave
superconductor.

\subsection{The similarities and discrepancies between $d$- and $s-$wave \
superconductors}

For magnetically mediated $d-$wave superconductivity, the role of phonons is
played by spin fluctuations. As we said, these excitations are propagating,
magnon-like modes below $T_{c}$ (more accurately, below the onset
temperature for the pseudogap), with the gap $\Delta _{s}$. This spin gap
obviously plays the same role as $\Delta _{p}$ for phonons, and hence we
expect that the spectral function should display \ a peak-dip-hump structure
as well. We will also demonstrate below that for the observables such as the
DOS, Raman intensity and the optical conductivity, which measure the
response averaged over the Fermi surface, the angular dependence of the $d-$%
wave gap $\Delta (\theta )\propto \cos \left( 2\theta \right) $ softens the
singularities, but does not wash them out over a finite frequency range.
Indeed, we find that the positions of the singularities are not determined
by some averaged gap amplitude but by the maximum value of the $d-$wave gap, 
$\Delta (0)=\Delta $. 

Despite similarities, the feedback effects for phonon-mediated $s-$wave
superconductors, and magnetically mediated $d-$wave superconductors are not
equivalent as we now demonstrate. The point is that for $s-$wave
superconductors, the exchange process shown in Fig.\ref{Figure1}a is not the
only possible source for the fermionic decay: there exists another process,
shown in Fig.\ref{Figure1}b, in which a fermion decays into three other
fermions. This process is due to a residual four-fermion interaction~\cite
{varma,coffey}. One can easily make sure that this second process also gives
rise to the fermionic decay when the external $\omega $ exceeds a minimum
energy of $3\Delta $, necessary to pull all three intermediate particles out
of the condensate of Cooper pairs. At the threshold, the fermionic spectral
function is non-analytic, much like at $\Delta +\Delta _{p}$. This implies
that in $s$-wave superconductors, there are two physically distinct
singularities, at $\Delta +\Delta _{p}$ and at $3\Delta $, which come from 
{\it different} processes and therefore are independent of each other. Which
of the two threshold frequencies is larger depends on the strength of the
coupling and on the shape of the phonon density of states. At weak coupling, 
$\Delta _{p}$ is exponentially larger than $\Delta $, hence $3\Delta $
threshold comes first. At strong coupling, $\Delta _{s}$ and $\Delta $ are
comparable, but calculations within Eliashberg formalism show that for real
materials ( e.g. for lead or niobium) still, 
$3\Delta <\Delta +\Delta _{p}$.~\cite
{mahan}. This result is fully consistent with the photoemission data for these
 materials~\cite{C2000}.

For magnetically mediated $d$-wave superconductors the situation is
different.  In the one-band model for cuprates, which we adopt, the
underlying interaction is solely a Hubbard-type four-fermion interaction.
The introduction of a spin fluctuation as an extra degree of freedom is just
a way to account for the fact that there exists a particular interaction
channel, where the effective interaction between fermions is the strongest
due to a closeness to a magnetic instability. This implies that the
propagator of spin fluctuations by itself is made out of particle-hole
bubbles like those in Fig.\ref{Figure1}b. Then, to the lowest order in the
interaction, the fermionic self-energy is given by the diagram in Fig.\ref
{Figure1}b. Higher-order terms convert a particle-hole bubble in Fig.\ref
{Figure1}b. into a wiggly line, and transform this diagram into the one in
Fig.\ref{Figure1}a. Clearly then, a simultaneous inclusion of both diagrams
would be a double counting, i.e., there is only a {\it single} process which
gives rise to the threshold in the fermionic self-energy.

Leaving a detailed justification of the spin-fermion model to the next
section, we merely notice here that the very fact that the diagram in Fig 
\ref{Figure1}b is a part of that in Fig.\ref{Figure1}a implies that the
development of a singularity in the spectral function at a frequency
different from $3\Delta$ cannot be due to effects outside the spin-fermion
model. Indeed, we will show that the model itself generates two
singularities, at $3\Delta$ and at $\Delta +\Delta _{s}<3\Delta $. The fact
that this is an internal effect, however, implies that $\Delta _{s}$ {\it %
depends} on $\Delta$. The experimental verification of this dependence 
can then be considered as a ``fingerprint'' of the spin-fluctuation
mechanism. Furthermore, as the singularities at $3\Delta $ and $\Delta
+\Delta _{s}$ are due to the same interaction, their relative intensity is
another gauge of the magnetic mechanism for the pairing. We will argue below
that some experiments on cuprates, particularly the measurements of optical
conductivity~\cite{CSB99}, allow one to detect both singularities, and that
their calculated relative intensity is consistent with the data.

\section{Spin-fermion model}

The point of departure for our analysis is the effective low-energy theory
for Hubbard-type lattice fermion models. As mentioned above, we assume {\it %
a'priori} that integrating out states with high fermionic energies in a
renormalization group sense, one obtains low-energy collective bosonic modes
only in the spin channel. In this situation, the low-energy theory should
include fermions, their collective bosonic spin excitations, and the
interaction between fermions and spins. This model is called a spin-fermion
model~\cite{Chubukov97}, and is described by 
\begin{eqnarray}
{\cal H} &=&\sum_{{\bf k},\alpha }{\bf v_{{\rm F}}\cdot }({\bf k}-{\bf k}%
_{F})c_{{\bf k},\alpha }^{\dagger }c_{{\bf k},\alpha }+\sum_{q}\chi
_{0}^{-1}({\bf q}){\bf S}_{{\bf q}}{\bf S}_{-{\bf q}}+  \nonumber \\
&&g\sum_{{\bf q,k},\alpha ,\beta }~c_{{\bf k+q},\alpha }^{\dagger }\,{\bf %
\sigma }_{\alpha ,\beta }\,c_{{\bf k},\beta }\cdot {\bf S}_{{\bf -q}}\,.
\label{intham}
\end{eqnarray}
Here $c_{{\bf k},\alpha }^{\dagger }$ is the fermionic creation operator for
an electron with crystal momentum ${\bf k}$ and spin $\alpha $, $\sigma _{i}$
are the Pauli matrices, and $g$ is the coupling constant which measures the
strength of the interaction between fermions and their collective bosonic
spin degrees of freedom, characterized by the spin-1 boson field, ${\bf S}_{%
{\bf q}}$. The latter are characterized by a bare spin susceptibility $\chi
_{0}({\bf q})=\chi _{0}\xi ^{2}/(1+({\bf q}-{\bf Q})^{2}\xi ^{2})$, where $%
\xi $ is the magnetic correlation length.

The relevant topological variable in the theory is the shape of the Fermi
surface. We assume that the Fermi-surface is hole-like, i.e., it is centered
at $(\pi ,\pi )$ rather than at $(0,0)$. This Fermi surface is consistent
with the photoemission measurements for ${\rm Bi}2212$, at least at and
below optimal doping. Luttinger theorem implies that this Fermi surface
necessary contains hot spots -- the points at the Fermi surface separated by
the antiferromagnetic momentum ${\bf Q}$. In ${\rm Bi}2212$, these hot spots
are located near $(0,\pi )$ and symmetry related points\cite
{Norman97,shennat}. The exact location of hot spots is however not essential
for our calculations. It is only important that hot spots do exist, are not
located close to the nodes of the superconducting gap, and that the Fermi
velocities at ${\bf k}_{{\rm hs}}$ and ${\bf k}_{{\rm hs}}+{\bf Q}$ are not
antiparallel to each other. 
We also assume that $\omega_0$ is smaller then $\epsilon_{0,\pi}$, i.e., the
van-Hove singularity of the electronic dispersion at $k =(0,\pi)$ is
irrelevant for our analysis.

Observe also that our bare $\chi _{0}({\bf q})$ does not depend on
frequency. In general, the integration over high energy fermions may give
rise to some frequency dependence of $\chi _{0} ({\bf q}, \omega)$. However,
as $\chi_0 ({\bf q}, \omega)$ comes  from fermions with $\omega \sim E_{{\rm %
F}}$, its frequency dependence  holds in powers of $\omega/E_{{\rm F}})^{2}$%
.  We will see that this frequency dependence can be safely neglected as it
is completely overshadowed by the $i\omega g^2 \chi_0/v^2_F $ term which
comes from low-energy fermions.  The presence of hot spots is essential in
this regard because a spin fluctuation with a momentum near ${\bf Q}$ can
decay into fermions at or near hot spots. By virtue of energy conservation
this process involves only low-energy fermions and therefore is fully
determined within the model.

This evolution of the bosonic dynamics from propagating to relaxational is
the key element which distinguishes between spin-mediated and
phonon-mediated superconductivities. For phonon superconductors, the bosonic
self-energy due to spin-fermion interaction  also contains a linear in $%
\omega$ term. However, this term has an extra relative smallness in $u/v_F
\propto m/M$, where $u$ is the sound velocity, $m$ is the electron mass, and 
$M$ is the mass of an ion. Due to this  extra smallness, the linear in $%
\omega$ term in the phonon propagator becomes relevant only at extremely low
frequencies, unessential for superconductivity. At $\omega \sim \Delta$, the
bare $\omega^2$ term dominates, i.e., the phonon propagator preserves its
input form. Alternatively speaking, the renormalization of the phonon
propagator by fermions is a minor effect while the same renormalization of
the spin propagator dominates the physics below $E_F$.

\subsection{Computational technique}

The input parameters in Eq. (\ref{intham}) are the coupling constant $g$,
the spin correlation length, $\xi $, the Fermi velocity $v_{F}$ (which we
assume to depend weakly on the position on the Fermi surface), and the
overall factor $\chi _{0}$. The latter, however, can be absorbed into the
renormalization of the coupling constant ${\bar{g}}=g^{2}\chi _{0}$, and
should not be counted as an extra variable. Out of these parameters 
one can construct a dimensionless ratio $\lambda =3{\bar{g}}/4\pi v_{F}\xi
^{-1}$ and an overall energy scale ${\bar{\omega}}=9{\bar{g}}/16\pi $ (the
factors $3/4\pi $ \ and $9/16\pi $ are introduced for further convenience).
All physical quantities discussed below can be expressed in terms of these
two parameters (and the angle between ${\bf v}_{{\bf k}_{{\rm hs}}}$ and $%
{\bf v}_{{\bf k}_{{\rm hs}}{\bf +Q}}$, which does not enter the theory in
any significant manner as long as ${\bf v}_{{\bf k}_{{\rm hs}}}$ and ${\bf v}%
_{{\bf k}_{{\rm hs}}{\bf +Q}}$ are not antiparallel to each other). One can
easily make sure that in two dimensions, a formal perturbation expansion
holds in powers of $\lambda $. The limit $\lambda \ll 1$ is perturbative and
is probably applicable only to strongly overdoped cuprates . The situation
in optimally doped and underdoped cuprates most likely corresponds to a
strong coupling, $\lambda \geq 1$. The most direct experimental indication
for this is the absence of a sharp quasiparticle peak in the normal state
ARPES data in materials with doping concentration equal to or below the
optimal one~ \cite{Norman97,shennat}.

At strong coupling, a conventional perturbation theory does not work, but we
found earlier that a controllable expansion is still possible if one
formally treats the number of hot spots in the Brillouin zone $N=8$ as a
large number.\cite{ac,acf,acs} The justification and a detailed description
of this procedure is beyond the scope of the present paper. We refer the
reader to the original publications, and quote here only the result: near
hot spots, one can obtain a set of coupled integral equations for three
complex variables: the anomalous vertex $\Phi _{{\bf k}}(\omega )\approx
\Phi _{{\bf k}_{{\rm hs}}}(\omega )$ subject to the $d$-wave constraint $%
\Phi _{{\bf k}}(\omega )=-\Phi _{{\bf k+Q}}(\omega )$, the fermionic
self-energy $\Sigma _{{\bf k}}(\omega )\approx \Sigma _{{\bf k}_{{\rm hs}%
}}(\omega )$, and the spin polarization operator $\Pi _{{\bf Q}}(\omega )$.
The anomalous vertex and the fermionic self-energy are related to \ the
normal and anomalous Green's functions as 
\begin{eqnarray}
G_{{\bf k}}(\omega ) &=&\frac{\Sigma _{{\bf k}}(\omega )+\varepsilon _{{\bf k%
}}}{\Sigma _{{\bf k}}^{2}(\omega )-\Phi _{{\bf k}}^{2}(\omega )-\varepsilon
_{{\bf k}}^{2}}, \\
F_{{\bf k}}(\omega ) &=&\frac{\Phi _{{\bf k}}(\omega )}{\Sigma _{{\bf k}%
}^{2}(\omega )-\Phi _{{\bf k}}^{2}(\omega )-\varepsilon _{{\bf k}}^{2}},
\label{GF}
\end{eqnarray}
and the polarization operator is related to the fully renormalized spin
susceptibility as 
\begin{equation}
\chi ({\bf q},\omega )=\frac{\chi _{0}\xi ^{2}}{1+({\bf q}-{\bf Q})^{2}\xi
^{2}-\Pi _{{\bf q}}(\omega )/\omega _{{\rm sf}}}.  \label{chif}
\end{equation}
We normalized $\Pi _{{\bf q}}(\omega )$ such that in the normal state $\Pi _{%
{\bf Q}}(\omega )=i\omega $ (see below). This normalization implies that $%
\omega _{{\rm sf}}={\bar{\omega}}/4\lambda ^{2}$

In Matsubara frequencies the set of the three equations has the form 
\begin{eqnarray}
\Phi _{m} &=&~\frac{\pi T}{2}\sum_{n}\frac{\Phi _{n}}{\sqrt{\Phi
_{n}^{2}+\Sigma _{n}^{2}}}~\left( \frac{{\bar{\omega}}}{\omega _{{\rm sf}%
}+\Pi _{n-m}}\right) ^{1/2}  \label{setphi} \\
\Sigma _{m} &=&\omega _{m}+\frac{\pi T}{2}\sum_{n}\frac{\Sigma _{n}}{\sqrt{%
\Phi _{n}^{2}+\Sigma _{n}^{2}}}~\left( \frac{{\bar{\omega}}}{\omega _{{\rm sf%
}}+\Pi _{n-m}}\right) ^{1/2}  \label{setsigma} \\
\Pi _{m} &=&\pi T~\sum_{n}\left( 1-\frac{\Sigma _{n}\Sigma _{n+m}+\Phi
_{n}\Phi _{n+m}}{\sqrt{\Phi _{n}^{2}+\Sigma _{n}^{2}}~\sqrt{\Phi
_{n+m}^{2}+\Sigma _{n+m}^{2}}}\right) .  \label{setpi}
\end{eqnarray}
Here, {\ }$\Phi _{m}=\Phi _{{\bf k}_{{\rm hs}}}(\omega _{m})$ and $\Sigma
_{m}=\Sigma _{{\bf k}_{{\rm hs}}}(\omega _{m})$ with fermionic Matsubara
frequency $\omega _{m}=\left( 2m+1\right) \pi T$ and $\Pi _{m}=\Pi _{{\bf Q}%
}(\omega _{m})$ with \ bosonic Matsubara frequency $\omega _{m}=2m\pi T$,
respectively. The first two equations are similar to the Eliashberg
equations for conventional superconductors. The presence of the third
coupled equation for $\Pi $ is peculiar to the spin-fluctuation scenario,
and reflects the fact that the spin dynamics is made out of fermions. As in
a conventional Eliashberg formalism, the superconducting gap $\Delta $ at $%
T=0$ is defined as a solution of $\Sigma (\omega )=\Phi (\omega )$, after
analytical continuation to the real axis.

The region around a hot spot where these equations are valid (i.e., the
``size'' of a hot spot) depends on frequency and is given by $|{\bf k}-{\bf k%
}_{{\rm hs}}|\sim \xi ^{-1}(1+|\omega |/\omega _{{\rm sf}})^{1/2}$~\cite
{acf,acs}. We will see that at strong coupling ($\lambda \geq 1$), the
superconducting gap $\Delta \sim \omega _{{\rm sf}}\lambda ^{2}$. Hence for
frequencies comparable to or larger than $\Delta $, typical $|{\bf k}-{\bf k}%
_{{\rm hs}}|\sim \lambda \xi ^{-1}\sim k_{F}({\bar{g}}/E_{F})$. In practice, 
${\bar{g}}$ is comparable to $E_F$ (in the RPA approximation for an
effective one-band Hubbard model for $CuO_2$, ${\bar g} \approx U \sim 2-3 eV
$, while $E_F$ is comparable to a bandwidth which has the same order of
magnitude). In this situation, the self energy and the anomalous vertex at
the hot spots are characteristic for the behavior of these functions in a
substantial portion of the Fermi surface, leading to an effective momentum
independence of the fermionic dynamics away from the nodes of the gap. Of
course, near zone diagonals, there is a different physics at low
frequencies, associated with the fact that the superconducting gap vanishes
for momenta along the diagonals. Below we show that the physics close to the
nodes is universally determined by the vanishing superconducting gap, and
therefore insensitive to strong coupling effects which bear fingerprints of
the pairing mechanism. For these reasons we will mostly concentrate our
analysis to describe peculiarities of the $d$-wave state at frequencies $%
\omega \geq \Delta _{{\rm \max }}$.

\section{The spin polarization operator}

\label{spo} Since our goal is to find the ``fingerprints'' of spin
excitations in the fermionic variables, we first discuss the general form of
the spin polarization operator. We show that in the normal state (ignoring
pseudogap effects), gapless fermions cause a purely diffusive spin dynamics.
However, in a $d-$wave superconducting state, a gap in the single particle \
dynamics gives rise to ''particle like'' propagating magnons.

\subsection{Normal state spin dynamics}

In the normal state $(\Phi (\omega )=0)$ the polarization operator can be
computed explicitly even without the knowledge of the precise form of $%
\Sigma (\omega )$. Indeed, from (\ref{setpi}) we immediately obtain 
\begin{equation}
\Pi _{m}=\pi T\sum_{n}\left( 1-\mbox{sign}(\Sigma _{n})\mbox{sign}(\Sigma
_{n+m})\right)  \nonumber
\end{equation}
The result depends only on the sign of the self-energy, but not on its
amplitude. As $\mbox{sign}(\Sigma _{n})=\mbox{sign}(2n+1)$, the computation
is elementary, and we obtain that for any coupling strength 
\begin{equation}
\Pi _{m}=|\omega _{m}|.  \label{normal-solution_pi}
\end{equation}
On the real axis this translates into a purely diffusive and overdamped spin
dynamics with\ $\Pi (\omega )=i\omega $. This result holds as long as the
linearization of the fermionic dispersion near the Fermi surface remains
valid, i.e., up to energies comparable to the Fermi energy. The linearity of 
$\Pi (\omega )$ and thus the fact that spin excitations in the normal state
can propagate only in a diffusive way is a direct consequence of the
presence of hot spots.

\subsection{Spin dynamics in the superconducting state}

The opening of the superconducting gap changes this picture. Now
quasiparticles near hot spots are gapped, so a spin fluctuation can decay
into a particle-hole pair only when it can pull two particles out of the
condensate of Cooper pairs. This implies that the decay \ into particle hole
excitations is only possible if the external frequency is larger than $%
2\Delta $. At smaller frequencies, $\Pi ^{\prime \prime }(\omega )=0$ for $%
T=0$. This result also readily follows from Eq.(\ref{setpi}). The
Kramers-Kronig relation $\Pi ^{\prime }(\omega )=(2/\pi )\int_{0}^{\infty
}\Pi ^{\prime \prime }(x)/(x^{2}-\omega ^{2})$ \ then implies that due to
the drop in $\Pi ^{\prime \prime }(\omega )$, the spin polarization operator
in a superconductor acquires a real part, which at low $\omega $ is
quadratic in frequency: $\Pi (\omega )\propto \omega ^{2}/\Delta $. An
essential point for our consideration is that $\Pi (\omega =0)=0$ for any $%
\Sigma (\omega )$ and $\Phi (\omega )$. This physically implies that the
development of the gap does not change the magnetic correlation length, a
result which becomes evident if one notices that $d$-wave pairing involves
fermions from opposite sublattices. 

Substituting the result $\Pi (\omega )\propto \omega ^{2}/\Delta $ into Eq.(%
\ref{chif}), we find that at low energies, spin excitations in a $d$-wave
superconductor are propagating, gaped magnon-like excitations: 
\begin{equation}
\chi ({\bf q},\omega )\propto \frac{1}{\Delta _{s}^{2}+c_{s}^{2}({\bf q}-%
{\bf Q})^{2}-\omega ^{2}}.  \label{chil}
\end{equation}
The magnon gap, $\Delta _{s}\sim (\Delta \omega _{{\rm sf}})^{1/2}$, and the
magnon velocity, $c_{s}^{2}\sim v_{{\rm F}}^{2}\Delta /{\bar{g}}$, are
entirely determined by the dynamics of the fermionic degrees of freedom.

Eq. (\ref{chil}) is meaningful only if $\Delta _{s}\leq \Delta $, i.e., $%
\omega _{{\rm sf}}\leq \Delta $. Otherwise the use of the quadratic form for 
$\Pi (\omega )$ is not justified. To find how $\Delta $ depends on the
coupling constant, one needs to carefully analyze the full set Eqn.(\ref
{setphi}-\ref{setpi}). This analysis is rather involved~\cite{acf,acs}, and
is not directly related to the goal of this paper. We skip the details and
just quote the result. It turns out that at strong coupling, $\lambda \leq 1$%
, i.e. for optimally and underdoped cuprates, the condition $\Delta >\omega
_{{\rm sf}}$ is satisfied as the gap scales with ${\bar{\omega}}$ and
saturates at $\Delta \approx 0.35{\bar{\omega}} = 0.06 {\bar g}$ at $\lambda
=\infty$, when $\omega_{sf} \propto \lambda^{-2}=0$. In this situation, the
spin excitations in a superconductor are propagating, particle-like modes
with the gap $\Delta_s$. However, in distinction to phonons, these
propagating magnons get their identity from a strong coupling feedback
effect in the superconducting state.

At weak coupling, the superconducting problem is of BCS type, and $\Delta
\sim \omega _{{\rm sf}}\exp \left( {-\lambda }^{-1}\right) \ll \omega _{{\rm %
sf}}$. This result is intuitively obvious as $\omega _{{\rm sf}}$ plays the
role of the Debye frequency in the sense that the bosonic mode which
mediates pairing decreases at frequencies above $\omega _{{\rm sf}}$. We see
that at weak coupling, the quadratic approximation for $\Pi (\omega )$ does
not lead to a pole in $\chi ({\bf Q},\omega )$. Still, the pole in $\chi
^{\prime \prime }({\bf Q},\omega )$ does exist also at weak coupling as we
now demonstrate. Indeed, consider $\Pi ^{\prime \prime }(\omega )$ at $%
\omega \approx 2 \Delta$. One can easily make sure that at this $\omega$,
one can simultaneously set both fermionic frequencies in the bubble to be
close to $\Delta $, and get a strong singularity due to vanishing of $\sqrt{%
\Sigma ^{2}-\Phi ^{2}}$ for both fermions. Substituting $\Sigma ^{2}(\omega
)-\Phi ^{2}(\omega )\propto \omega -\Delta $ into (\ref{setpi}) and using
the spectral representation, we obtain for $\omega =2\Delta +\epsilon $ 
\begin{equation}
\Pi ^{\prime \prime }(\omega )\propto \int_{0}^{\epsilon }\frac{dx}{%
(x(\epsilon -x))^{1/2}}.
\end{equation}
Evaluating the integral, we find that $\Pi ^{\prime \prime }$ undergoes a
finite jump at $\omega =2\Delta $. By Kramers-Kronig relation, this jump
gives rise to a logarithmic singularity in $\Pi ^{\prime }(\omega )$ at $%
\omega =2\Delta $: 
\begin{equation}
\Pi ^{\prime }(\omega )=\frac{2}{\pi }~\int_{2\Delta }^{\infty }dx\frac{\Pi
^{\prime \prime }(x)}{x^{2}-\omega ^{2}}\propto \Delta \log \frac{2\Delta }{%
|\omega -2\Delta |}.
\end{equation}
\begin{figure}[tbp]
\begin{center}
\epsfxsize=3.0in \epsfysize=1.6in
\epsffile{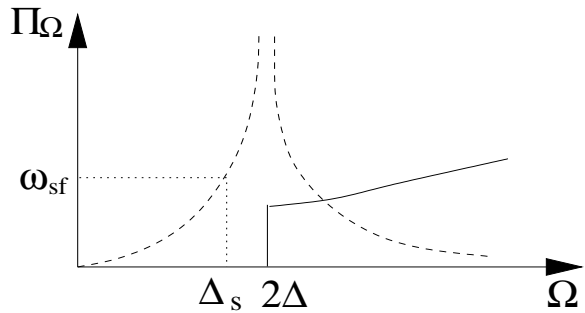}
\end{center}
\caption{Schematic behavior of the real (dashed line) and imaginary (solid
line) parts of the particle hole bubble in the superconducting state. Due to
the discontinuous behavior of $\Pi ^{\prime \prime }\left( \protect\omega
\right) $ at $\protect\omega =2\Delta $, the real part $\Pi ^{\prime }(%
\protect\omega )$ is logarithmically divergent at $2\Delta $. For small $%
\protect\omega $, the real part behaves like $\protect\omega ^{2}/\Delta $.}
\label{Figure3}
\end{figure}
The behavior of $\Pi ^{\prime }(\omega )$ and $\Pi ^{\prime \prime }(\omega
) $ is schematically shown in Fig.\ref{Figure3}. The fact that $\Pi ^{\prime
}(\omega )$ diverges logarithmically at $2\Delta $ implies that no matter
how small $\Delta /\omega _{sf}$ is, $\chi ({\bf Q},\omega )$ has a pole at $%
\Delta _{s}<2\Delta $, when $\Pi ^{\prime \prime }(\omega )$ is still zero.
Simple estimates show that for weak coupling, where $\omega _{{\rm sf}}\gg
\Delta $, the singularity occurs at $\Delta _{s}=2\Delta (1-Z_{s})$ where $%
Z_{s}\propto e^{-\omega _{sf}/(2\Delta )}$ is also the spectral weight of
the resonance peak in this limit.

We see therefore that the resonance in the spin susceptibility exists both
at weak and at strong coupling. At strong coupling, the resonance frequency
is $\Delta _{s}\sim \Delta /\lambda \ll \Delta$, i.e., the resonance occurs
in the frequency range where spin excitations behave as propagating
magnon-like excitations. At weak coupling, the resonance occurs very near $%
2\Delta $ due to the logarithmic singularity in $\Pi ^{\prime }(\omega )$.
In practice, however, the resonance at weak coupling can hardly be observed
because the residue of the peak in the spin susceptibility $Z_{s}$ is
exponentially small.

\begin{figure}[tbp]
\begin{center}
\epsfxsize=3.0in \epsfysize=1.6in
\epsffile{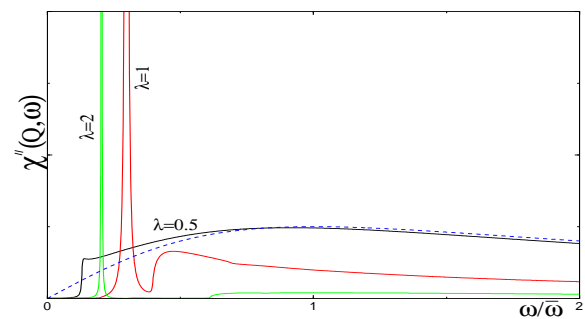}
\end{center}
\caption{Imaginary part of the dynamical spin susceptibility in the
superconducting state for $T\ll \Delta $ and $\protect\lambda =0.5, 1,2$,
determined from the full solution of the Eliashberg equations. Dashed line -
the normal state result shown for comparison. Observe that the resonance
peak gets sharper when it moves away from $2\Delta$.}
\label{Figure4}
\end{figure}

Fig.\ref{Figure4} shows our results for   $\chi ({\bf Q},\omega )$ obtained
from the full solution of the set of three coupled equations at $T\approx 0$
and $\lambda =0.5,1,2$. We clearly see that for $\lambda \geq 1$, the spin
susceptibility has a sharp peak at $\omega =\Delta _{s}$. The peak gets
sharper when it moves away from $2\Delta $. At the same time, for $\lambda
=0.5$, which models weak coupling, the peak is very weak and is washed out
by a small thermal damping. In this case, $\chi ^{\prime \prime }$ only
displays a discontinuity at $2\Delta $.

Before we proceed with the analysis of fermionic properties, we show that
the spin resonance does not exist for $s-$wave superconductors. In the
latter case, the spin polarization operator is given by almost the same
expression as in (\ref{setpi}), but with a different sign for $\Phi _{n}\Phi
_{n+m}$ term. The sign difference comes from the fact that the two fermions
in the spin polarization bubble differ in momentum by ${\bf Q}$, and the $d$%
-wave gap changes sign under ${\bf k}\rightarrow {\bf k}+{\bf Q}$.

One can immediately check using (\ref{setpi}) that for a different sign of
the anomalous term, $\Pi ^{\prime \prime }$ does not possess a jump at $%
2\Delta $ (the singular contributions from $\Sigma _{n}\Sigma _{n+m}$ and $%
\Phi _{n}\Phi _{n+m}$ cancel each other). Then $\Pi ^{\prime }(\omega )$
does not diverge at $2\Delta $, and hence there is no resonance at weak
coupling. Still, however, one could naively expect the resonance at strong
coupling as at small frequencies $\Pi ^{\prime }(\omega )$ is quadratic in $%
\omega $ simply by virtue of the existence of the threshold for $\Pi
^{\prime \prime }$. It turns out, however, that the resonance in the case of
isotropic $s$-wave pairing is precluded by the fact that $\Pi (\omega =0)<0 $
(recall that in case of $d$-wave pairing, $\Pi (\omega =0)=0$). This
negative term overshadows the positive $\omega ^{2}$ term in $\Pi (\omega )$
such that for all frequencies below $2\Delta $, $\Pi (\omega )<0$~\cite{oleg}%
. That $\Pi (\omega =0)<0$ in $s-$wave superconductors can be easily
explained. Indeed, a negative $\Pi (0)$ implies that the spin correlation
length decreases as the system becomes superconducting. This is exactly what
one should expect as $s$-wave pairing involves fermions both from opposite
and from the same sublattice. The pairing of fermions from the same
sublattice into a spin-singlet state obviously reduces the antiferromagnetic
correlation length.

\subsection{Comparison with experiments}

\begin{figure}[tbp]
\begin{center}
\epsfxsize=3.0in \epsfysize=1.6in
\epsffile{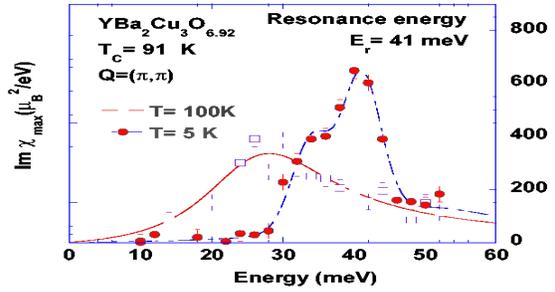}
\end{center}
\caption{Inelastic neutron scattering intensity for momentum ${\bf Q=}\left( 
\protect\pi ,\protect\pi \right) $ as function of frequency for {\rm YBa}$%
_{2}{\rm Cu}_{3}{\rm O}_{6.9}$. Data from Ref.~\protect\cite{Bourges99}.}
\label{Figure_resonance_exp}
\end{figure}

In Fig.\ref{Figure_resonance_exp} we show the representative experimental
result for $\chi ^{\prime \prime }({\bf Q},\omega )$ for optimally doped $%
{\rm YBi}_{2}{\rm Cu}_{3}{\rm O}_{6.9}$\cite{Bourges99}. We clearly see a
resonance peak at $\omega \approx 41$ ${\rm meV}$. A very similar result has
been recently obtained for ${\rm Bi}2212$\cite{neutrons2}. In the last case,
the resonance frequency is $43$ ${\rm meV}$. With underdoping, the measured
resonance frequency goes down~\cite{neutrons,dai}. In strongly underdoped $%
{\rm YBi}_{2}{\rm Cu}_{3}{\rm O}_{6.6}$, it is approximately $25$ ${\rm meV}$%
\cite{neutrons}. The very existence of the peak and the downturn
renormalization of its position with underdoping agree with the theory.
Furthermore, the measured $\chi ^{\prime \prime }({\bf Q},\omega )$ displays
an extra feature at $60-80$ ${\rm meV} $~\cite{dai,Bourges99}, which can be
possibly explained as a $2\Delta $ effect.

The full analysis of the resonance peak requires more care as (i) the peak
is only observed in two-layer materials, and only in the odd channel, (ii)
the momentum dispersion of the peak is more complex than that for magnons~ 
\cite{Bourges00}, (iii) the peak broadens with underdoping~\cite
{neutrons,dai}, (iv) in underdoped materials, the peak emerges at the onset
of the pseudogap, and only sharpens at $T_{c}$~\cite{dai,Bourges99}. All
these features can be explained within the spin-fermion model as well. The
discussion of these effects, however, is beyond the scope of the present
paper. For our present purposes, it is essential that the resonance peak has
been observed experimentally, and that its frequency for optimally doped
materials is around $40$ ${\rm meV}$.

We now proceed to the detailed analysis how the spin resonance at $\Delta
_{s}$ affects fermionic observables. In each instance, we first briefly
discuss the results for a $d-$wave gas, and then focus on the strong
coupling limit.

\section{``Fingerprints'' of the spin resonance in fermionic variables}

\subsection{The spectral function}

We first consider the spectral function $A_{{\bf k}}(\omega )=(1/\pi )|G_{%
{\bf k}}^{\prime \prime }(\omega )|$. In the superconducting state, for
quasiparticles near the Fermi surface 
\begin{equation}
A_{{\bf k}}(\omega >0)=\frac{1}{\pi }~{\rm Im}\left[ \frac{\Sigma (\theta
,\omega )+\varepsilon _{{\bf k}}}{\Sigma ^{2}(\theta ,\omega )-\Phi
^{2}(\theta ,\omega )-\varepsilon _{{\bf k}}^{2}}\right] ,  \label{sf}
\end{equation}
where, we remind, $\Sigma (\theta ,\omega )$ is the fermionic self-energy
(which includes a bare $\omega $ term), and $\Phi (\theta ,\omega )$ is a $%
d- $wave anomalous vertex. Both can be taken at the Fermi surface, and
generally depend on the direction of ${\bf k}_{{\rm F}}$, which we labeled
as $\theta $. Also, by definition, $A_{{\bf k}}(-\omega )=A_{{\bf k}}(\omega
)$.

\subsubsection{The $d$-wave gas}

In a Fermi gas with $d$-wave pairing, $\Sigma (\theta ,\omega )=\omega $,
and $\Phi (\theta ,\omega )=\Delta (\theta )\propto \cos \left( 2\theta
\right) $. The spectral function then has a $\delta -$functional peak at $%
\omega =(\Delta ^{2}(\theta )+\varepsilon _{{\bf k}}^{2})^{1/2}$. It is
obvious but essential for comparison with the strong coupling case that the
peak disperses with ${\bf k}$ and far away from the Fermi surface recovers
the normal state dispersion.

\subsubsection{Strong coupling behavior}

Here we consider the spectral function, $A_{{\bf k}}(\omega )$, for fermions
located near hot spots. As we discussed, for those fermions, one can obtain
a closed set of equations (Eqs.\ref{setphi}-\ref{setpi}) which determine $A_{%
{\bf k}}(\omega )$. Since spin fluctuations in a superconducting phase are
propagating quasiparticles, the effect of the spin scattering on fermions
near hot spots should be exactly the same as the effect of phonon scattering
in an $s$-wave superconductor, i.e., in addition to a peak at $\omega
=\Delta $, the spectral function $A_{{\bf k}_{{\rm hs}}}(\omega )=A(\omega )$
should possess a singularity at $\omega =\omega _{0}=\Delta +\Delta _{s}$.

The behavior of $A(\omega )$ near the singularity is very robust and can be
obtained even without a precise knowledge of the frequency dependence of $%
\Sigma (\omega )$ and $\Phi (\omega )$. Our reasoning here parallels the one
for $\Pi (\omega )$: all we need to know is that near $\omega =\Delta $, $%
\Sigma ^{2}(\omega )-\Phi ^{2}(\omega )\propto \omega -\Delta $.
Substituting this form into (\ref{setpi}) and using the spectral
representation, we obtain for $\omega =\omega _{0}+\epsilon $ 
\begin{equation}
\Sigma ^{\prime \prime }(\omega )\propto \int_{0}^{\epsilon }~\frac{dx}{%
(x(\epsilon -x))^{1/2}}
\end{equation}
Evaluating the integral, we find that $\Sigma ^{\prime \prime }$ undergoes a
finite jump at $\omega =\omega _{0}$. By Kramers-Kronig relation, this jump
gives rise to a logarithmic divergence of $\Sigma ^{\prime }$. Exactly the
same singular behavior holds for the anomalous vertex $\Phi (\omega )$, with
exactly the same prefactor in front of the logarithm. The last result
implies that $\Sigma (\omega )-\Phi (\omega )$ is non-singular at $\omega
=\omega _{0}$. Substituting these results into (\ref{sf}), we find that the
spectral function $A(\omega )$ emerges at $\omega >\omega _{0}$ as $1/\log
^{2}(\omega -\omega _{0})$, i.e., almost discontinuously. Obviously, at a
small but finite $T$, the spectral function should have a dip very near $%
\omega =\omega _{0}$, and a hump at a somewhat higher frequency. 
\begin{figure}[tbp]
\begin{center}
\epsfxsize=3.0in \epsfysize=4.8in
\epsffile{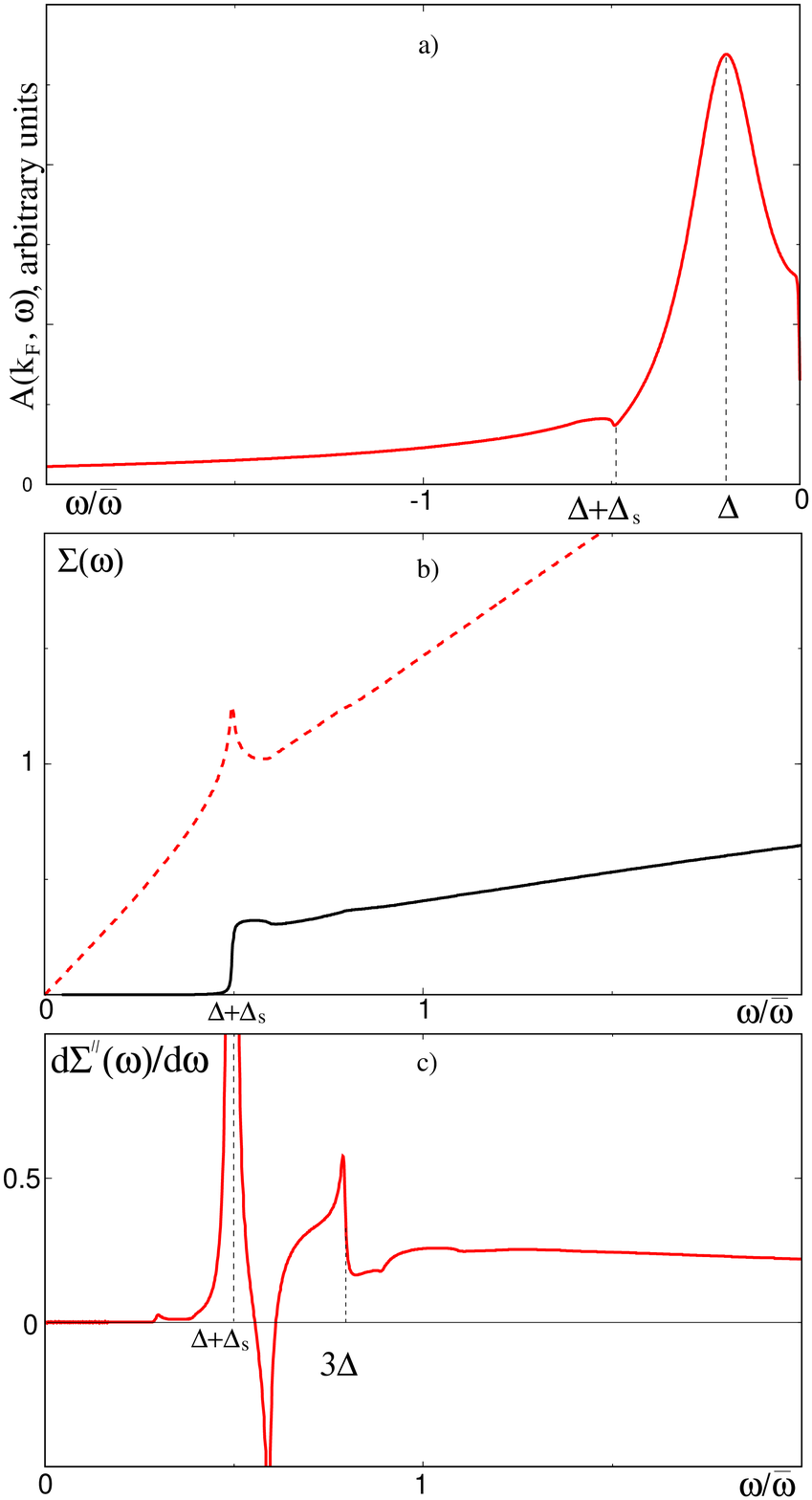}
\end{center}
\caption{(a) The quasiparticle spectral function determined by solving the
coupled Eliashberg equations. The peak-dip-hump structure of $A(\protect%
\omega )$ is clearly visible. (b) The real (dashed line) and imaginary
(solid line) parts of the electronic self-energy. (c) The derivative of the
imaginary part of the self-energy over $\protect\omega $. The singularities
at both $\Delta +\Delta _{s}$ and at $3\Delta $ are clearly seen.}
\label{Figure_specdens}
\end{figure}

In Fig.\ref{Figure_specdens} we present $\Sigma (\omega)$ and $A(\omega)$
from a solution of the set of three coupled Eliashberg equations at $%
T\approx 0$. This solution is consistent with our analytical estimate. We
clearly see that the fermionic spectral function has a peak-dip-hump
structure, and the peak-dip distance exactly equals $\Delta _{s}$. We also
see in Fig.\ref{Figure_specdens} that the spectral function is nonanalytic
at $\omega =3\Delta $. As we discussed, this nonanalyticity is peculiar to
the spin-fermion model, and is due to the nonanalyticity of the dynamical
spin susceptibility at $\omega =2\Delta $. 

\begin{figure}[tbp]
\begin{center}
\epsfxsize=3.0in \epsfysize=1.6in
\epsffile{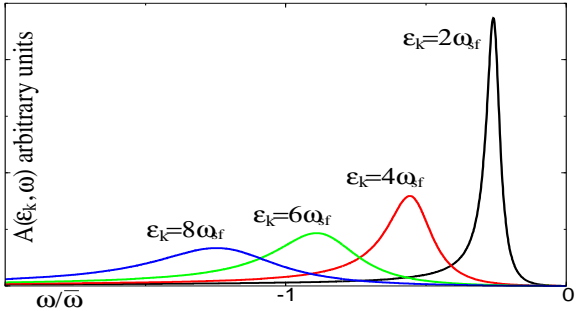}
\end{center}
\caption{The normal state spectral function, relevant for the high frequency
behavior in the superconducting state. Note the absence of a quasiparticle
peak. This is the consequence of the proximity to an antiferromagnetic
quantum critical point. }
\label{Figure_NFL}
\end{figure}

Another ``fingerprint'' of the spin-fluctuation scattering can be observed
by studying the evolution of the spectral function as one moves away from
the Fermi surface. The argument here goes as follows: at strong coupling,
where $\Delta \geq \omega _{{\rm sf}}$, probing the fermionic spectral
function at frequencies progressively larger than $\Delta $, one eventually
probes the normal state fermionic self-energy at $\omega \gg \omega _{{\rm sf%
}}$. Due to strong spin-fermion interaction, this self-energy is large.
Indeed, the solution of Eq.(\ref{setpi}) with $\Phi =0$ and $\Pi (\omega
)=i\omega $ yields at $T=0$\cite{Chubukov97} 
\begin{equation}
\Sigma (\omega )=\omega \left( 1+\frac{2\lambda }{1+(1-\frac{i|\omega |}{%
\omega _{{\rm sf}}})^{1/2}}\right)  \label{normal-solution_sigma}
\end{equation}
For $\omega \gg \omega _{sf}$, $\ \Sigma (\omega )$ becomes 
\begin{equation}
\Sigma (\omega )=\omega (1+(i{\bar{\omega}}/|\omega |)^{1/2})
\end{equation}
Substituting this form into the fermionic propagator, we find that up to $%
\omega \sim {\bar{\omega}}$, the spectral function in the normal state does
not have a quasiparticle peak at $\omega =\varepsilon _{{\bf k}}$. Instead,
it only displays a broad maximum at $\omega =\varepsilon _{{\bf k}}^{2}/{%
\bar{\omega}}$. Alternatively speaking, at $\omega _{{\rm sf}}<\omega <{\bar{%
\omega}}$, the spectral function in the normal state displays a non-Fermi
liquid behavior with no quasiparticle peak (see Fig.~\ref{Figure_NFL}). This
particular non-Fermi liquid behavior ($\Sigma (\omega )\propto \sqrt{\omega }
$) is associated with the closeness to a quantum phase transition into an
antiferromagnetic state. Indeed, at the transition point, $\omega _{{\rm sf}%
}=0$, and hence $\Sigma (\omega )\propto \sqrt{\omega }$ extends down to a
zero frequency.

The absence of the quasiparticle peak in the normal state implies that the
sharp quasiparticle peak which we found at $\omega =\Delta $ for momenta at
the Fermi, cannot simply disperse with ${\bf k}$, as it does for
noninteracting fermions with a $d$-wave gap. Specifically, the quasiparticle
peak cannot move further in energy than $\Delta +\Delta _{s}$ as at larger
frequencies, the spin scattering becomes possible, and the fermionic
spectral function should display roughly the same non-Fermi-liquid behavior
as in the normal state.

\begin{figure}[tbp]
\begin{center}
\epsfxsize=3.0in \epsfysize=1.7in
\epsffile{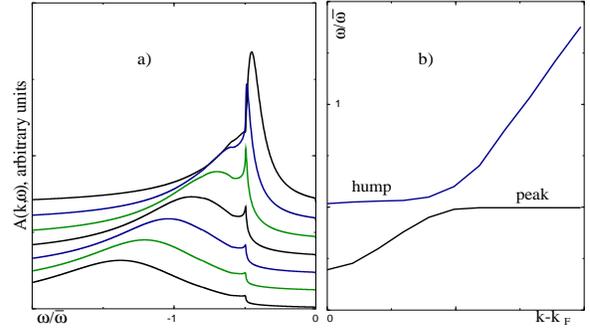}
\end{center}
\caption{a) Frequency dependence of the spectral function in the
superconducting state for different $\protect\epsilon_k$. The curve at the
bottom has a highest $\protect\epsilon_k$. No coherent quasiparticle peak
occurs for energies larger than $\Delta + \Delta_s$. Instead, the spectral
function displays a broad maximum, similar to that in the normal state.}
\label{Figure_spec_sc_mom}
\end{figure}

In Fig.\ref{Figure_spec_sc_mom}a we present successive plots for the
spectral function as the momentum moves away from the Fermi surface. We see
exactly the behavior we just described: the quasiparticle peak cannot move
further than $\Delta + \Delta_s$. Instead, when ${\bf k}-{\bf k}_{{\rm F}}$
increases, it gets pinned at $\Delta + \Delta_s$ and gradually looses its
spectral weight. At the same time, the hump disperses with ${\bf k}$ and for
frequencies larger than $\Delta + \Delta_s$ gradually transforms into a
broad maximum at $\omega =\varepsilon _{{\bf k}}^{2}/{\bar{\omega}}$. The
positions of the peak and the dip versus ${\bf k}-{\bf k}_{{\rm F}}$ are
presented in Fig.\ref{Figure_spec_sc_mom}b

\subsubsection{Comparison with the experiments}

\begin{figure}[tbp]
\begin{center}
\epsfxsize=2.0in \epsfysize=2.2in
\epsffile{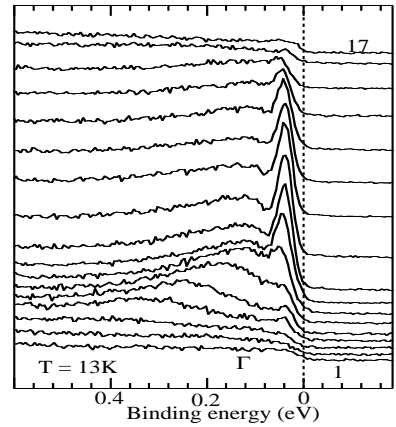}
\end{center}
\caption{ARPES spectrum \ for near optimally doped {\rm Bi}$2212$ for
momenta close to the hot spots. Data from Ref.[6].}
\label{Figure_ARPES_exp}
\end{figure}

The quasiparticle spectral function at various momenta is measured in angle
resolved photoemission (ARPES) experiments. In a sudden approximation (an
electron, hit by light, leaves the crystal without further interactions with
other electrons and without paying attention to selection rules for the
optical transition to its final state), the photoemission intensity is given
by $I_{{\bf k}}(\omega )=A_{{\bf k}}(\omega )n_{F}(\omega )$ where $n_{F}$
is the Fermi function. ARPES data of Ref.\cite{Norman97} for near optimally
doped, $T_{c}=87{\rm K}$ in ${\rm Bi}2212$ for momenta near a hot spot are
presented in Fig.\ref{Figure_ARPES_exp}. We clearly see that the intensity
displays a peak/dip/hump structure. A sharp peak is located at $\sim 40{\rm %
meV}$, and the dip is at $80{\rm meV}$.

\begin{figure}[tbp]
\begin{center}
\epsfxsize=3.0in \epsfysize=2.4in
\epsffile{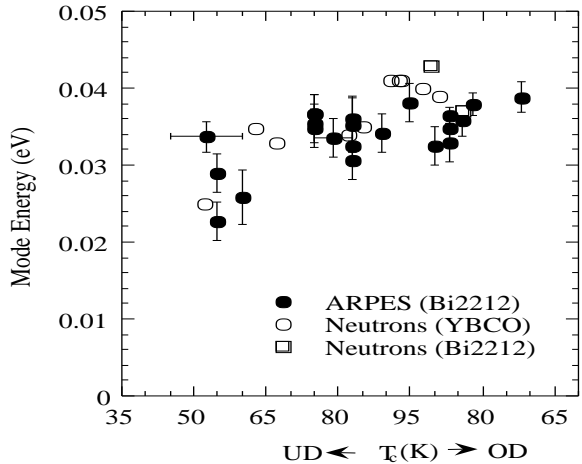}
\end{center}
\caption{The experimental peak-dip distance at various doping concentrations
vs $\Delta_s$ extracted from neutron measurements. Data from Ref.[30]}
\label{Figure_Arpes_peak_dip}
\end{figure}

The experimental peak-dip distance is $42{\rm meV}$ ~\cite{Norman97}. The
neutron scattering data~\cite{neutrons2} on ${\rm Bi}2212$ with nearly the
same $T_{c}=91{\rm K}$ yielded $\Delta _{s}=43{\rm meV}$ which is in
excellent agreement with the ARPES data. Furthermore, with underdoping, the
peak-dip energy difference decreases and, up to error bars, remains equal to 
$\Delta _{s}$. This behavior is illustrated in Fig.\ref
{Figure_Arpes_peak_dip}.

\begin{figure}[tbp]
\begin{center}
\epsfxsize=2.0in \epsfysize=2.0in
\epsffile{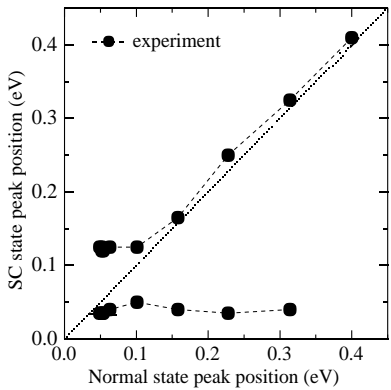}
\end{center}
\caption{The experimental peak and hump positions with the deviation from
the Fermi surface. We clearly see that the hump disperses with ${\bf k}-{\bf %
k}_{{\rm F}}$ and eventually recovers the position of the broad maximum in
the normal state, while the peak position changes little with the deviation
from ${\bf k}_{{\rm F}}$. Data from Ref.[6].}
\label{Figure_ARPES_awaykF}
\end{figure}

Finally, in Fig.\ref{Figure_ARPES_awaykF} we present experimental results
for the variation of the peak and hump positions with the deviation from the
Fermi surface. We clearly see that the hump disperses with ${\bf k}-{\bf k}_{%
{\rm F}}$ and eventually recovers the position of the broad maximum in the
normal state. At the same time, the peak shows little dispersion, and does
not move further in energy than $\Delta + \Delta_s$. Instead, the amplitude
of the peak just dies off as ${\bf k}$ moves away from ${\bf k}_{{\rm F}}$.
This behavior is fully consistent with the theory.

We regard the presence of the dip at $\Delta +\Delta _{s}$, and the absence
of the dispersion of the quasiparticle peak are two major ``fingerprints''
of strong spin-fluctuation scattering in the spectral density of cuprate
superconductors.

\subsection{The density of states}

The quasiparticle density of states, $N(\omega )$, is the momentum integral
of the spectral function: 
\begin{equation}
N(\omega )=\int \frac{d^{2}{\bf k}}{4\pi^2}~ A_{{\bf k}}(\omega ).
\end{equation}
Substituting $A_{{\bf k}}(\omega )$ from Eq.(\ref{sf}) and integrating over $%
\varepsilon _{{\bf k}}$, one obtains 
\begin{equation}
N(\omega )\propto {\rm Im}~\int_{0}^{2\pi }d\theta ~\frac{\Sigma (\theta
,\omega )}{(\Phi ^{2}(\theta ,\omega )-\Sigma ^{2}(\theta ,\omega ))^{1/2}},
\label{eq}
\end{equation}
where, we remind, $\theta $ is the angle along the Fermi surface. As before,
we first consider $N(\omega )$ in a $d$-wave gas, and then discuss strong
coupling effects.

\subsubsection{Density of states in a $d$-wave gas}

\begin{figure}[t]
\begin{center}
\leavevmode
\epsfxsize=3.3in 
\epsfysize=1.8in 
\epsffile{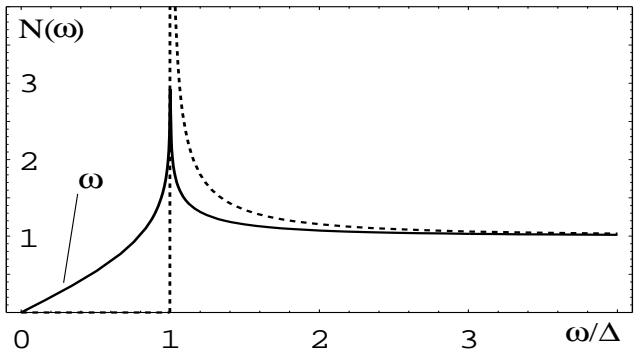}
\end{center}
\caption{Density of states of a noninteracting Fermi gas with $d$-wave gap
(solid line) and with $s$-wave gap (dashed line).}
\label{Figure_DOS_gas}
\end{figure}

Consider for simplicity a circular Fermi surface for which $\Delta _{{\bf k}%
}=\Delta \cos \left( 2\theta \right) $. Integrating in (\ref{eq}) over $%
\theta$ we obtain~\cite{maki} 
\begin{eqnarray}
N(\omega ) &=&{\rm Re}~\left[ \frac{\omega }{2\pi }\int_{0}^{2\pi }\frac{%
d\theta }{\sqrt{\omega ^{2}-\Delta ^{2}\cos ^{2}(2\theta )}}\right] 
\nonumber \\
&=&\frac{2}{\pi }\left\{ 
\begin{array}{ll}
K(\Delta /\omega ) & \mbox{for $\omega >\Delta $} \\ 
(\omega /\Delta )K(\omega /\Delta ) & \mbox{for $\omega <\Delta $}
\end{array}
\right. ,  \label{sin-gas}
\end{eqnarray}
where $K(x)$ is the elliptic integral of first kind. We see that $N(\omega
)\sim \omega $ for $\omega \ll \Delta $ and diverges logarithmically as $%
(1/\pi )\ln (8\Delta /|\Delta -\omega |)$ for $\omega \approx \Delta $. At
larger frequencies, $N(\omega )$ gradually decreases to a frequency
independent, normal state value of the DOS, which we normalized to $1$. The
plot of $N(\omega )$ in a $d-$wave BCS superconductor is presented in Fig.%
\ref{Figure_DOS_gas} 
For comparison, in an $s$-wave superconductor, the DOS vanishes at $\omega
<\Delta $ and diverges as $(\omega -\Delta )^{-1/2}$ at $\omega \geq \Delta $%
. We see that a $d$-wave superconductor is different in that (i) the DOS is
finite down to the smallest frequencies, and (ii) the singularity at $\omega
=\Delta $ is weaker (logarithmic). Still, however, $N(\omega )$ is singular
only at a frequency which equals to the largest value of the $d-$wave gap.
This illustrates a point we made earlier that the angular dependence of the $%
d$-wave gap reduces the strength of the singularity, but does not wash it
out over a finite frequency range.

\subsubsection{Density of states at strong coupling}

We first show that the linear behavior of $N(\omega )$ at low frequencies
and the logarithmic divergence at $\omega =\Delta $, observed in a gas, are
in fact quite general and are present in an arbitrary $d-$wave
superconductor. Indeed, at low frequencies ${\rm Re}\Sigma (\theta ,\omega
)\propto \omega $, ${\rm Re}\Phi (\theta ,\omega )\propto (\theta -\theta _{%
{\rm node}})$ where $\theta _{{\rm node}}$ $=\frac{\pi }{4},\frac{3\pi }{4}%
...$ are the positions of the node of the $d$-wave gap. Substituting these
forms into (\ref{eq}) and integrating over $\theta $ we obtain $N(\omega
)\propto \omega $. \ Similarly, expanding $\Sigma ^{2}-\Phi ^{2}$ near a hot
spot, where, at least at strong coupling, the $d-$wave gap is at maximum~ 
\cite{acf}, we obtain $\Sigma ^{2}(\theta ,\omega )-\Phi ^{2}(\theta ,\omega
)\propto (\omega -\Delta )+B\widetilde{\theta }^{2}$, where $\widetilde{%
\theta }=\theta -\theta _{{\rm hs}}$, and $B>0$. \ Here, $\theta _{{\rm hs}}$
is the position of a hot spot on the Fermi surface. The integration over $%
\widetilde{\theta }$ then yields 
\begin{equation}
N(\omega )\propto {\rm Re}\int \frac{d\widetilde{\theta }}{\sqrt{B\widetilde{%
\theta }^{2}+(\Omega -\Delta )}}\approx -\frac{\log |\Omega -\Delta |}{\sqrt{%
B}}.  \label{sin-strong}
\end{equation}
This result implies that at strong coupling, the DOS in a $d-$wave
superconductor still has a logarithmic singularity at $\omega =\Delta $,
although the overall factor for the singular piece depends on the strength
of the interaction.

We see therefore that the behavior of the density of states in a $d$-wave
superconductors for $\omega \leq \Delta $ is quite robust against strong
coupling phenomena. This nice universality on the other hand makes $N(\omega
\leq \Delta )$ not very sensitive to a specific mechanism for the pairing.

\begin{figure}[t]
\begin{center}
\leavevmode
\epsfxsize=3.3in 
\epsfysize=1.8in 
\epsffile{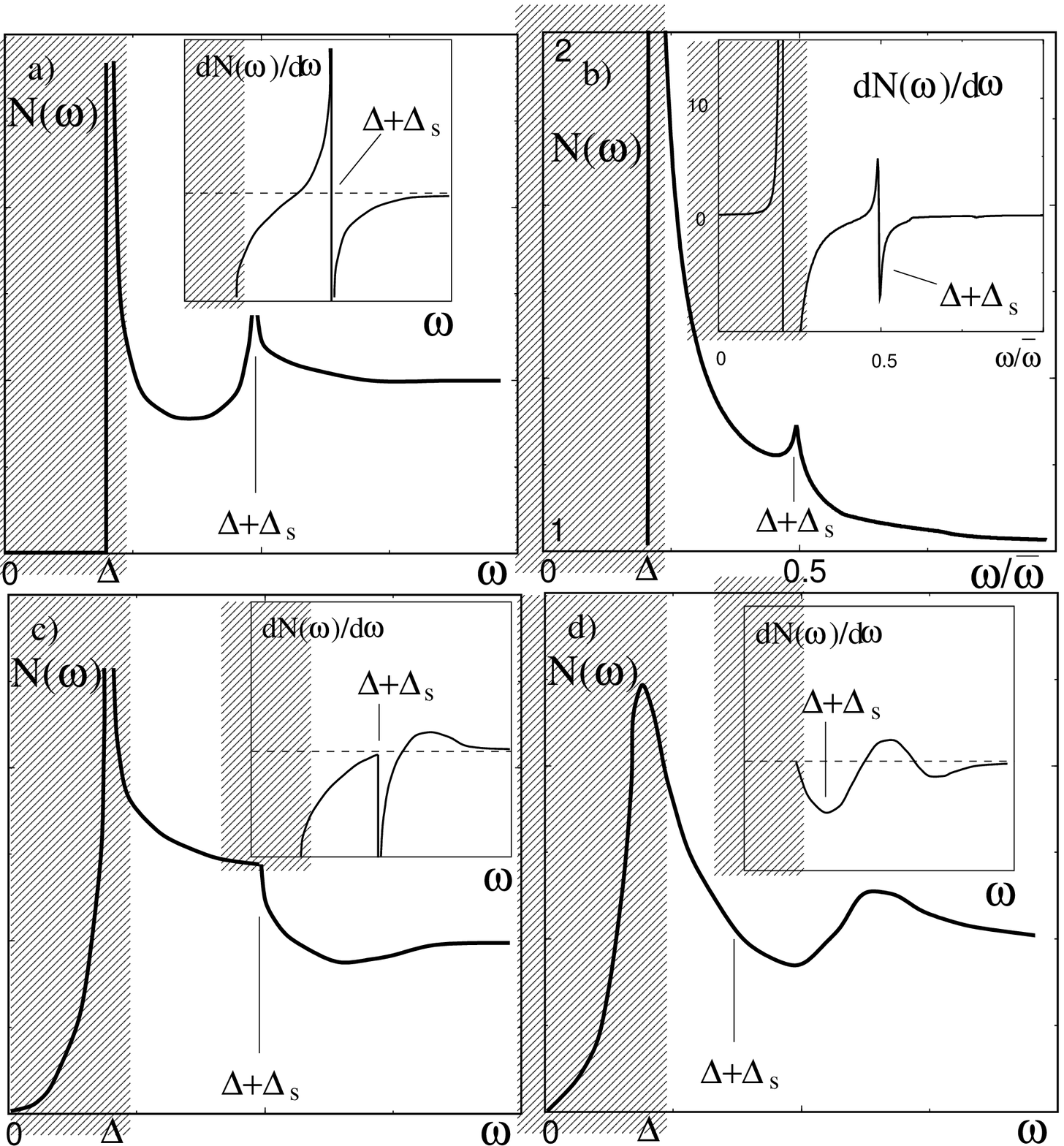}
\end{center}
\caption{(a) The behavior of the SIN tunneling conductance (i.e., DOS) in
strongly coupled $d$-wave superconductor. Main pictures - $N(\protect\omega)$%
, insets - $dN(\protect\omega)/d\protect\omega$. (a) The schematic behavior
of the DOS for a flat gap. (b) The solution of the the set of the
Eliashberg-type equations for flat gap. The shaded regions are the ones in
which the flat gap approximation is incorrect as the physics is dominated by
nodal quasiparticles. (c) The schematic behavior of $N(\protect\omega)$ for
the quadratic variation of the gap near its maxima. (d) The expected
behavior of the DOS in a real situation when singularities are softened out
by finite $T$ or impurity scattering. The position of $\Delta + \Delta_s$
roughly corresponds to the minimum of $dN(\protect\omega)/d \protect\omega$.}
\label{Figure_DOS_schematic}
\end{figure}

We now demonstrate that at strong coupling, the DOS possesses extra peak-dip
features, associated with the singularities in $\Sigma (\omega )$ and $\Phi
(\omega )$ at $\omega =\omega _{0}=\Delta +\Delta _{s}$. The analytical
consideration proceeds as follows~\cite{acsin}. Consider first a case when
the gap is totally flat near a hot spot. At $\omega =\omega _{0}$, both $%
\Sigma (\omega )$ and $\Phi (\omega )$ diverge logarithmically. Substituting
these forms into (\ref{eq}), we immediately obtain that $N(\omega )$ has a
logarithmic singularity: 
\begin{equation}
N_{{\rm sing}}(\omega )\propto \left( \log {\frac{1}{|\omega -\omega _{0}|}}%
\right) ^{1/2}.
\end{equation}
This singularity gives rise to a strong divergence of $dN(\omega )/d\omega $
at $\omega =\omega _{0}$. This behavior is schematically shown in Fig. (\ref
{Figure_DOS_schematic})a. In part (b) of this figure we present the result
for $N(\omega )$ obtained by the solution of Eliashberg-type Eqs.\ref{setphi}%
-\ref{setpi}. A small but finite temperature was used to smear out
divergences. We recall that the Eliashberg set does not include the angular
dependence of the gap near hot spots, and hence the result for the DOS
should be compared with Fig. (\ref{Figure_DOS_schematic})a. We clearly see
that $N(\omega )$ has a second peak at $\omega =\omega _{0}$. This peak
strongly affects the frequency derivative of $N(\omega )$ which has a
predicted singular behavior near $\omega _{0}$.

Note also that a relatively small magnitude of the singularity in $N(\omega) 
$ is a consequence of the linearization of the fermionic dispersion near the
Fermi surface. For actual $\epsilon_k$ chosen to fit ARPES data~\cite{mike},
the nonlinearities in the fermionic dispersion occur at energies comparable
to $\omega_0$. This is due to the fact that hot spots are located close to $%
(0,\pi)$ and related points at which the Fermi velocity vanishes. As a
consequence, the momentum integration of the spectral function should have
less drastic smearing effect than in our calculations, and the frequency
dependence of $N(\omega)$ should more resemble that of $A(\omega)$ for
momenta where the gap is at maximum. 
For momentum dependent gap, the behavior of fermions near hot spots is the
same as when the gap is flat, but now $\omega _{0}$ depends on $\theta $ as
both $\Delta $ and $\Delta _{s}$ vary as one moves away from a hot spot. The
variation of $\Delta $ is obvious, the variation of $\Delta _{s}$ is due to
the fact that this frequency scales as $\Delta ^{1/2}$. Since both $\Delta $
and $\Delta _{s}$ are maximal at or very near a hot spot, we can write quite
generally 
\begin{equation}
\omega_0 \rightarrow \omega_0 - a {\tilde \theta}^2.
\end{equation}
where $a >0$. The singular pieces of the self-energy and the anomalous
vertex then behave as $\log (\omega _{0}-\omega -a\widetilde{\theta }%
^{2})^{-1}$. Substituting these forms into (\ref{eq}) and using the fact
that $\Sigma (\omega )-\Phi (\omega )\approx const$ at $\omega \approx
\omega _{0}$, we obtain 
\begin{equation}
N_{{\rm sing}}(\omega ) \propto {\rm Re}\int d\widetilde{\theta }\left[\log
(\omega _{0}-\omega -a\widetilde{\theta }^{2})^{-1}\right]^{-1/2}.
\end{equation}
A straightforward analysis of the integral shows that now $N(\omega )$ has a
one-sided nonanalyticity at $\omega =\omega _{0}$: 
\begin{equation}
N_{{\rm sing}}(\omega )=-B\Theta (\omega -\omega _{0})\!\!\!~\left( \frac{%
\omega -\omega _{0}}{|\log (\omega -\omega _{0})|}\right) ^{1/2},  \label{nd}
\end{equation}
where $B>0$, and $\Theta (x)=1$ for $x>0$, and $\Theta (x)=0$ for $x<0$.
This nonanalyticity gives rise to a cusp in $N(\omega)$ right above $%
\omega_0 $, and a one-sided square-root divergence of the frequency
derivative of the DOS. This behavior is shown schematically in Fig. (\ref
{Figure_DOS_schematic})c. Comparing this behavior with the one in Fig. (\ref
{Figure_DOS_schematic})a for a flat gap, we observe that the angular
dependence of the gap predominantly affects the form of $N(\omega)$ at $%
\omega \leq \omega_0$. At these frequencies, the angular variation of the
gap completely eliminates the singularity in $N(\omega)$. At the same time,
above $\omega_0$, the angular dependence of the gap softens the singularity,
but still, the DOS sharply drops above $\omega_0$ such that the derivative
of the DOS diverges at approaching $\omega_0$ from above. Alternatively
speaking, in a $d-$wave superconductor, the singularity in the DOS is
softened by the angular dependence of the gap, but it still holds at a
particular frequency related to the maximum value of the gap. This point is
essential as it enables us to read off the maximum gap value directly from
the experimental data without ''deconvolution'' \ of momentum averages.

For real materials, in which singularities are removed by e.g., impurity
scattering, $N(\omega )$ likely has a dip at $\omega \geq \omega _{0}$ and a
hump at a larger frequency. This is schematically shown in Fig. (\ref
{Figure_DOS_schematic})d. The location of $\omega_0$ is best described as a
point where the frequency derivative of the DOS passes through a minimum.

The singularity in $dN(\omega )/d\omega $ at $\omega _{0}$, and the dip-hump
structure of $N(\omega )$ at $\omega \geq \omega _{0}$ are another
``fingerprints'' of spin-fluctuation mechanism in the single particle
response.

\subsubsection{Comparison with the experiments}

\begin{figure}[t]
\begin{center}
\leavevmode
\epsfxsize=3.3in 
\epsfysize=1.8in 
\epsffile{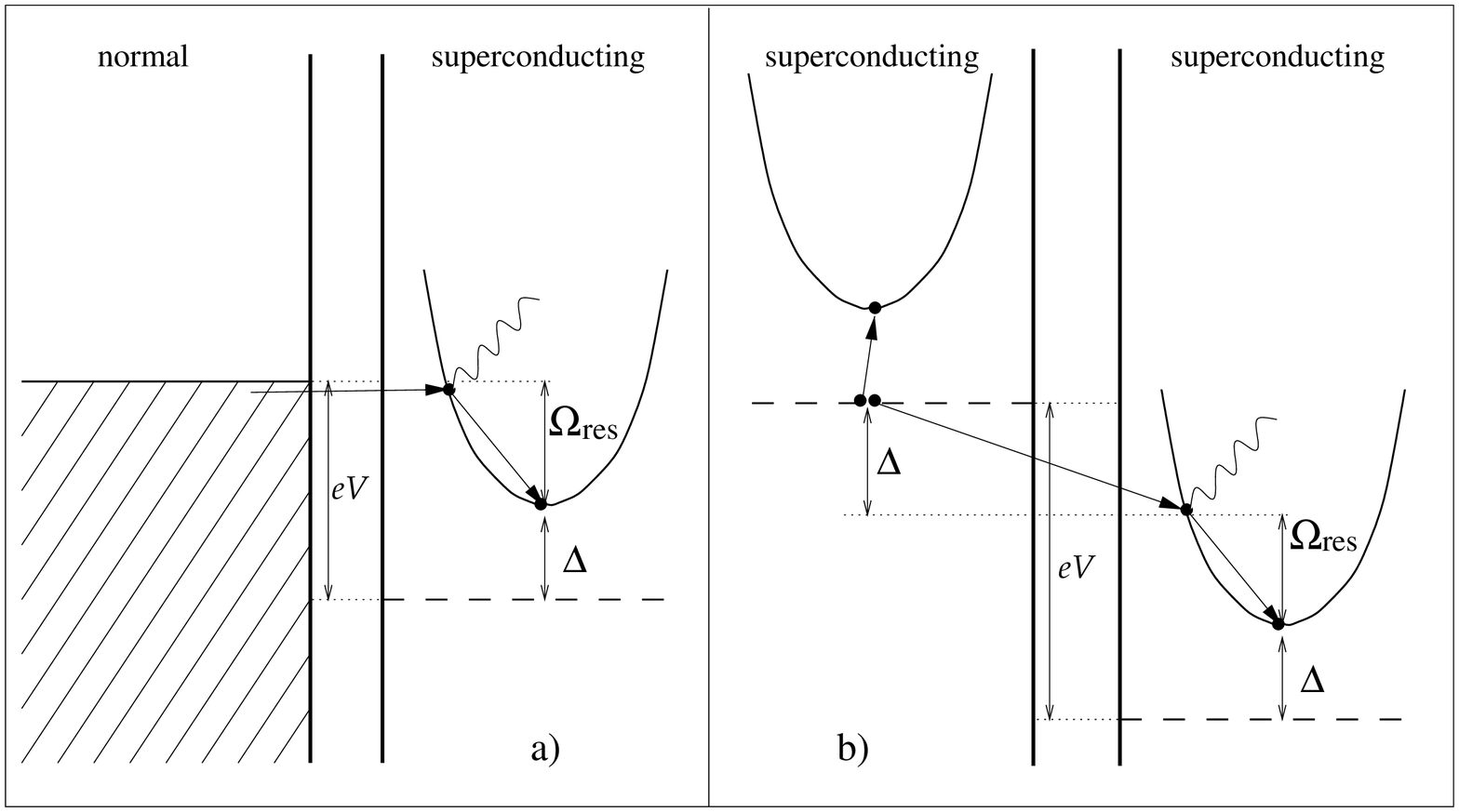}
\end{center}
\caption{The schematic diagram for the dip features in the SIN and SIS
tunneling conductances (figures a and b, respectively). For the SIN
tunneling, which measures fermionic DOS, the electron which tunnels from a
normal metal can emit a propagating magnons if the voltage $eV=\Delta
+\Delta _{s}$. After emitting a magnon, the electron falls to the bottom of
the band. This leads to a sharp reduction of the current and produces a drop
in $dI/dV$. For SIS tunneling, the physics is similar, but one first has to
break an electron pair, which costs energy $2\Delta $.}
\label{sins}
\end{figure}

\begin{figure}[tbp]
\begin{center}
\epsfxsize=3.0in \epsfysize=2.4in
\epsffile{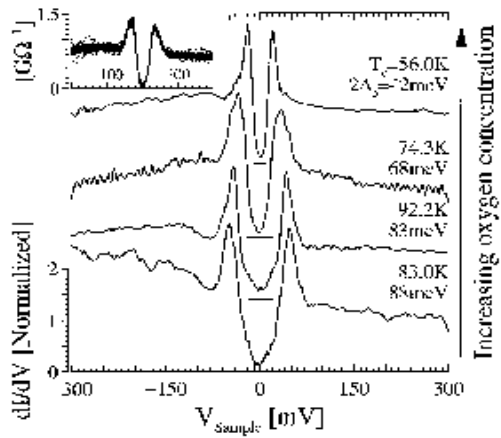}
\end{center}
\caption{The experimental result for the differential conductance through
the SIN tunneling junction. This conductance is proportional to the
quasiparticle DOS. The data are for $Bi2212$ and are taken from Ref.~ 
\protect\cite{fisher}.}
\label{sin}
\end{figure}

As we already mentioned, the fermionic DOS $N(\omega )$ is proportional to
the dynamical conductance $dI/dV$ through a superconductor-insulator-normal
metal (SIN) at $\omega =eV~$\cite{mahan}. The drop in the DOS at $\omega _{0}
$ can be reformulated in terms of SIN conductance as follows: if the voltage
for SIN tunneling is such that $eV=\omega _{0}$, then an electron which
tunnels from the normal metal, can emit a spin excitation and fall to the
bottom of the band loosing its group velocity. This obviously leads to a
sharp reduction of the current and produces a drop in $dI/dV$. This process
is shown schematically in Fig\ref{sins}~\cite{commnew}. 
The SIN tunneling experiments have been performed on ${\rm YBCO}$ and ${\rm %
Bi}2212$ materials~\cite{fisher}. We reproduce these data in Fig.~\ref{sin}.
Very similar results have been recently obtained by Pan et al~\cite{davis}. 

At low and moderate frequencies, the SIN conductance displays a behavior
which is generally expected in a $d-$wave superconductor, i.e., it is linear
in voltage for small voltages, and has a peak at $eV=\Delta $ where $\Delta $
is the maximum value of the $d-$wave gap~\cite{fisher,davis}. The value of $%
\Delta $ extracted from tunneling agrees well with the maximum value of the
gap extracted from ARPES measurements~\cite{Fedorov99,Kaminski00}. At
frequencies larger than $\Delta $, the measured SIN conductance clearly
displays an extra dip-hump feature which become visible at around optimal
doping, and grows in amplitude with underdoping~\cite{fisher}. At optimal
doping, the distance between the peak at $\Delta $ and the dip is around $40%
{\rm meV}$. This is consistent with $\Delta _{s}$ extracted from neutron
measurements. The doping dependence of the peak-dip distance, and the
behavior of $dN(\omega )/d\omega $ have not been studied in detail, to the
best of our knowledge. This analysis is clearly called for.

\subsection{SIS tunneling}

The measurements of the dynamical conductance $dI/dV$ through a
superconductor - insulator - superconductor (SIS) junction is another tool
to search for the fingerprints of the spin-fluctuation mechanism. The
conductance through this junction is the derivative over voltage of the
convolution of the two DOS~\cite{mahan}: $dI/dV\propto S(\omega )$ where 
\begin{equation}
S(\omega )=\int_{0}^{\omega }d\Omega N(\omega -\Omega )~\partial _{\Omega
}N(\Omega )  \label{sis}
\end{equation}

\subsubsection{SIS tunneling in a $d$-wave gas}

\begin{figure}[t]
\begin{center}
\leavevmode
\epsfxsize=3.0in 
\epsfysize=1.8in 
\epsffile{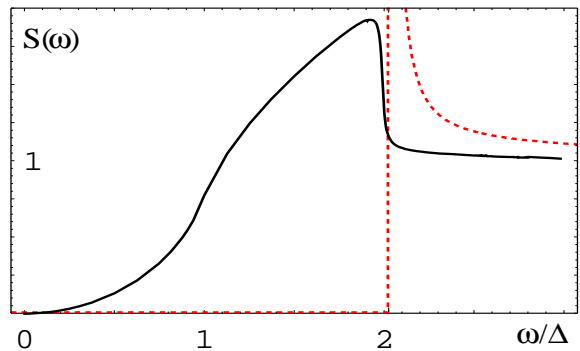}
\end{center}
\caption{The SIS tunneling conductance, $dI/dV$, in a $d-$wave BCS
superconductor. The dashed line shows $S(\protect\omega )$ for $s$-wave case
for comparison.}
\label{FIGURE_SIS}
\end{figure}

The DOS in a $d-$wave gas is given in Eq.(\ref{sin-gas}). Substituting this
form into Eq.(\ref{sis}) and integrating over frequency we obtain the result
presented in Fig.\ref{FIGURE_SIS}. At small $\omega $, $S(\omega )$ is
quadratic in frequency~\cite{maki}. This is an obvious consequence of the
fact that the DOS is linear in $\omega $. At $\omega =2\Delta $, $S(\omega )$
undergoes a finite jump. This jump is related to the fact that near $2\Delta 
$, the integral over the two DOS includes the region around $\Omega =\Delta $
where both $N(\Omega )$ and $N(\omega -\Omega )$ are logarithmically
singular, and $\partial _{\Omega }N(\Omega )$ diverges as $1/(\Omega -\Delta
)$. The singular contribution to $S(\omega )$ from this region can be
evaluated analytically and yields 
\begin{equation}
S(\omega )=-\frac{1}{\pi ^{2}}P\int_{-\infty }^{\infty }\frac{dx~\log |x|}{%
x+\omega -2\Delta }=-\frac{1}{2}\mbox{sign}(\omega -2\Delta )
\label{sis-gas}
\end{equation}
Observe that the amplitude of the jump in the SIS conductance is a universal
number which does not depend on the value of $\Delta $.

At larger frequencies, $S(\omega)$ continuously goes down and eventually
approaches a value of $S(\omega \rightarrow \infty)=1$.

\subsubsection{SIS tunneling at strong coupling}

As in the previous subsection, we first demonstrate that the quadratic
behavior at low frequencies and the discontinuity at $2\Delta $ survive at
arbitrary coupling. Indeed, the quadratic behavior at low $\omega $ is just
a consequence of the linearity of $N(\omega )$ at low frequencies. As shown
above, this linearity is a general property of a $d-$wave superconductor.
Similarly, the logarithmic divergence of the DOS at $\omega =\Delta $ causes
the discontinuity in the SIS conductance by the same reasons as in a $d-$%
wave gas.

\begin{figure}[t]
\begin{center}
\leavevmode
\epsfxsize=3.3in 
\epsfysize=1.8in 
\epsffile{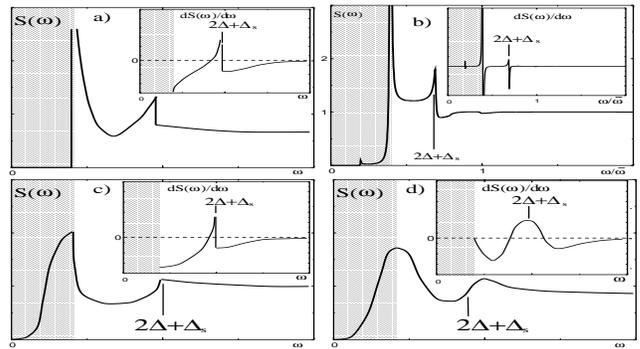}
\end{center}
\caption{ (a) The schematic behavior of the SIS tunneling conductance, $S(%
\protect\omega)$, in a strongly coupled $d$-wave superconductor. Main
pictures - $S(\protect\omega)$, insets - $dS(\protect\omega)/d\protect\omega$%
. (a) The schematic behavior of $S(\protect\omega)$ for a flat gap. (b) The
solution of the the set of the Eliashberg-type equations for flat gap using
the DOS from the previous subsection. The shaded regions are the ones in
which the flat gap approximation is incorrect as the physics is dominated by
nodal quasiparticles. (c) The schematic behavior of $S(\protect\omega)$ for
quadratic variation of the gap near its maxima. (d) The expected behavior of
the SIS conductance in a real situation when singularities are softened out
by finite $T$ or by impurity scattering. $2\Delta + \Delta_s$ roughly
corresponds to the maximum of $dS(\protect\omega)/d \protect\omega$.}
\label{FIGURE_SIS_calc}
\end{figure}

We next consider how the singularity in $\Sigma (\omega )$ at $\omega _{0}$
affects $S(\omega )$. From a physical perspective, we should expect a
singularity in $S(\omega )$ at $\omega =\Delta +\omega _{0}=2\Delta +\Delta
_{s}$. Indeed, to get a nonzero SIS conductance, one has to first break a
Cooper pair, which costs an energy of $2\Delta $. After a pair is broken,
one of the electrons becomes a quasiparticle in a superconductor and takes
the energy $\Delta $, while the other tunnels. If $eV=\Delta +\omega _{0}$,
the electron which tunnels through a barrier has energy $\omega _{0}$, and
can emit a spin excitation and fall to the bottom of the band (see Fig.~\ref
{sins}). This should produce a drop in $dI/dV$ by the same reasons as for
SIN tunneling. This behavior is schematically shown in Fig.~\ref
{FIGURE_SIS_calc}.

Consider this effect in more detail. We first notice that $\omega =\Delta
+\omega _{0}$ is special for Eq.(\ref{sis}) because both $dN(\Omega
)/d\Omega $ and $N(\omega -\Omega )$ diverge at the same energy, $\Omega
=\omega_0$. Substituting the general forms of $N(\omega )$ near $\omega
=\omega _{0}$ and $\omega =\Delta $, we obtain after simple manipulations
that for a flat gap, $S(\omega)$ has a one-sided divergence at $\omega =
\omega_0 + \Delta = 2 \Delta + \Delta_s$. 
\begin{equation}
S_{{\rm sing}} (\epsilon)\propto \frac{\Theta (-\epsilon )}{\sqrt{-\epsilon }%
}
\end{equation}
where $\epsilon =\omega -(\omega _{0}+\Delta)$. This obviously causes the
divergence of the frequency derivative of $S(\omega)$ (i.e., of $%
d^{2}I/dV^{2}$). This behavior is schematically shown in Fig. \ref
{FIGURE_SIS_calc}a. In Fig. \ref{FIGURE_SIS_calc}b we present the results
for $S(\omega )$ obtained by integrating theoretical $N(\omega )$ from the
previous subsection (see Fig.\ref{Figure_DOS_schematic}b). We clearly see
that $S(\omega )$ and its frequency derivative are singular at $\omega
=2\Delta + \Delta_s$, in agreement with the analytical prediction.

For quadratic variation of the gap near the maxima, the calculations similar
to those for the SIN tunneling yield that $S(\omega)$ is continuous through $%
2\Delta + \Delta_s$, but its frequency derivative diverges as 
\begin{equation}
\frac{dS(\omega )}{d\omega }\propto P\int_{0}^{\Delta }\frac{dx}{(x|\log
x|)^{1/2}\left( x-\epsilon \right) }~\sim \frac{\Theta (-\epsilon )}{%
|\epsilon \log |\epsilon ||^{1/2}},  \label{sis2}
\end{equation}
The singularity in the derivative implies that near $\epsilon =0$ 
\begin{equation}
S(\epsilon )=S(0)-C~\Theta (-\epsilon )\left( \frac{-\epsilon}{|\log
(-\epsilon )|}\right)^{1/2},
\end{equation}
where $C>0$. This behavior is schematically presented in Fig. \ref
{FIGURE_SIS_calc}d. We again see that the angular dependence of the gap
softens the strength of the singularity, but the singularity remains
confined to a single frequency $\omega =2\Delta +\Delta_s$.

In real materials, the singularity in $S(\omega )$ is softened and
transforms into a dip slightly below $2\Delta +\Delta_s$, and a hump at a
frequency larger than $2\Delta + \Delta_s $. The frequency $2\Delta +
\Delta_s$ roughly corresponds to a maximum of the frequency derivative of
the SIS conductance.

\subsubsection{comparison with experiments}

\begin{figure}[tbp]
\begin{center}
\epsfxsize=3.0in \epsfysize=3.0in
\epsffile{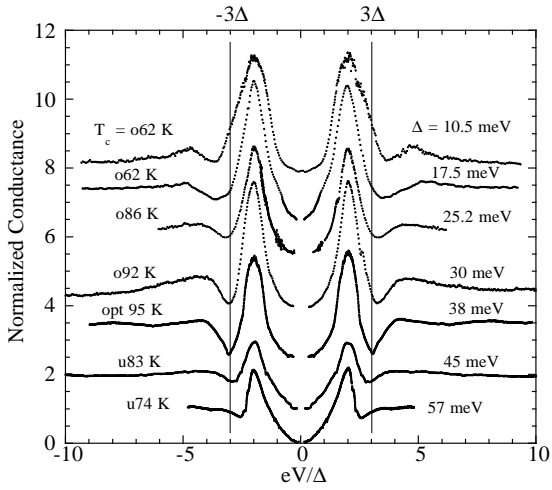}
\end{center}
\caption{SIS tunneling data from Ref.[11] for a set of ${\rm Bi}2212$
materials ranging from overdoped to underdoped.}
\label{Figure_SIS_Zasa}
\end{figure}

Recently, Zasadzinski {\em et al.}\cite{zasad} obtained and carefully
examined their SIS tunneling data for a set of ${\rm Bi}2212$ materials
ranging from overdoped to underdoped\cite{zasad}. Their data, presented in
Fig.\ref{Figure_SIS_Zasa}, clearly show that besides a peak at $2\Delta $,
the SIS conductance also has a dip and a hump at larger frequencies. The
distance between the peak and the dip ($\approx \Delta _{s}$ in our theory)
is close to $2\Delta $ in overdoped ${\rm Bi}2212$ materials, but goes down
with underdoping. Near optimal doping, this distance is around $40{\rm meV}$%
. For underdoped, $T_{c}=74{\rm K}$ material, the peak-dip distance is
reduced to about $30{\rm meV}$. 

These results are in qualitative and quantitative agreement with ARPES and
neutron scattering data, as well as with our theoretical estimates. The most
important aspect is that with underdoping, the experimentally measured
peak-dip distance progressively shifts down from $2\Delta $. This is the key
feature of the spin-fluctuation mechanism, We regard the experimental
verification of this feature in the SIS tunneling data is a strong argument
in favor of a magnetic scenario for superconductivity.

\subsection{Optical conductivity and Raman response}

Another observables sensitive to $\omega _{0}$ are the optical conductivity
and the Raman response. 
Both are proportional to the fully renormalized particle-hole polarization
bubble, but with different signs attributed to the bubble made of anomalous
propagators. Namely, after integrating in the particle-hole bubble over $%
\varepsilon _{{\bf k}}$, one obtains 
\begin{eqnarray}
R(\omega ) &=&{\rm Im}\int d\omega ^{\prime }d\theta V^{2}(\theta )\Pi
_{r}(\theta ,\omega ,\omega ^{\prime }) \\
\sigma (\omega ) &=&{\rm Im}\frac{i}{\omega +i\delta }~\int d\omega ^{\prime
}d\theta \Pi _{\sigma }(\theta ,\omega ,\omega ^{\prime })  \label{rs}
\end{eqnarray}
where $V(\theta )$ is a Raman vertex which depends on the scattering
geometry \cite{girsh}, and 
\begin{equation}
\Pi _{r,\sigma }(\theta ,\omega ,\omega ^{\prime })=~\frac{\Sigma _{+}\Sigma
_{-}+\alpha \Phi _{+}\Phi _{-}+D_{+}D_{-}}{D_{+}D_{-}(D_{+}+D_{-})}
\label{pi}
\end{equation}
Here $\alpha =-1$ for $\Pi _{r}$, and $\alpha =1$ for $\Pi _{\sigma }$.
Also, $\Sigma _{\pm }=\Sigma \left( \omega _{\pm }\right) $ and $\Phi _{\pm
}=\Phi \left( \omega _{\pm }\right) $ with $\omega _{\pm }=\omega ^{\prime
}\pm \omega /2$ as well as $D_{\pm }=(\Phi _{\pm }^{2}-\Sigma _{\pm
}^{2})^{1/2}$. Note, $\Sigma $ and $\Phi $ depend on $\omega $ and $\theta $.

\subsubsection{Optical and Raman response in a $d$-wave gas}

In a superconducting gas, the optical conductivity vanishes identically for
any nonzero frequency due to the absence of a physical scattering between
quasiparticles in a gas. The presence of a superconducting condensate,
however, gives rise to a $\delta $ functional term in $\sigma $ at $\omega
=0 $: $\sigma (\omega )=\pi \delta (\omega )\int d\theta d\omega ^{\prime
}\Pi _{\sigma }(\theta ,0,\omega ^{\prime })$. This behavior is typical for
any BCS superconductor~\cite{schrieffer},  and holds for both $s-$wave and $%
d-$wave superconductors. The behavior of $\sigma (\omega )$ in a
superconducting gas with impurities, causing inelastic scattering, is more
complex and has been discussed for a $d-$wave case in,  e.g., Ref.\cite
{Quinlan96}.

The form of the Raman intensity depends on the scattering geometry. For
mostly studied $B_{1g}$ scattering, the Raman vertex has the same angular
dependence as the $d$-wave gap, i.e., $V(\theta )\propto \cos \left( 2\theta
\right)$~\cite{girsh,Devereaux}. Straightforward computations then show that
at low frequencies, $R(\omega )\propto \omega ^{3}$\cite{Devereaux}. For a
constant $V(\theta )$, we would have $R(\omega )\propto \omega $.

Near $\omega =2\Delta $, the $B_{1g}$ Raman intensity is singular. For this
frequency, both $D_{+}$ and $D_{-}$ vanish at $\omega ^{\prime }=0$ and $%
\theta =0$. This causes the integral for $R(\omega )$ to be divergent. The
singular contribution to $R(\omega )$ can be obtained analytically by
expanding in the integrand to leading order in $\omega ^{\prime }$ and in $%
\theta $. Using the spectral representation, we then obtain, for $\omega
=2\Delta +\delta $\cite{girsh} 
\begin{eqnarray}
R(\omega ) &=&\int_{0}^{\epsilon }d\Omega \int d{\tilde \theta} ~\frac{1}{%
\sqrt{\Omega +a{\tilde \theta}^{2}}\sqrt{\delta -\Omega +a{\tilde \theta}^{2}%
}}  \nonumber \\
&&\frac{1}{(\sqrt{\Omega +a{\tilde \theta}^{2}}+\sqrt{\delta -\Omega + a{%
\tilde \theta}^{2}})}  \label{gi}
\end{eqnarray}
Where, as before, ${\tilde \theta} = \theta -\theta_{hs}$ For a flat band ($%
a=0$), $R(\omega) \propto Re [(\omega - 2\Delta)^{-1/2}]$. For $a \neq 0$,
i.e., for a quadratic variation of the gap near its maximum, the 2D
integration in (\ref{gi}) yields 
$R(\omega )\propto |\log \epsilon | $. At larger frequencies $R(\omega )$
gradually decreases.

\begin{figure}[tbp]
\begin{center}
\epsfxsize=2.8in \epsfysize=1.8in
\epsffile{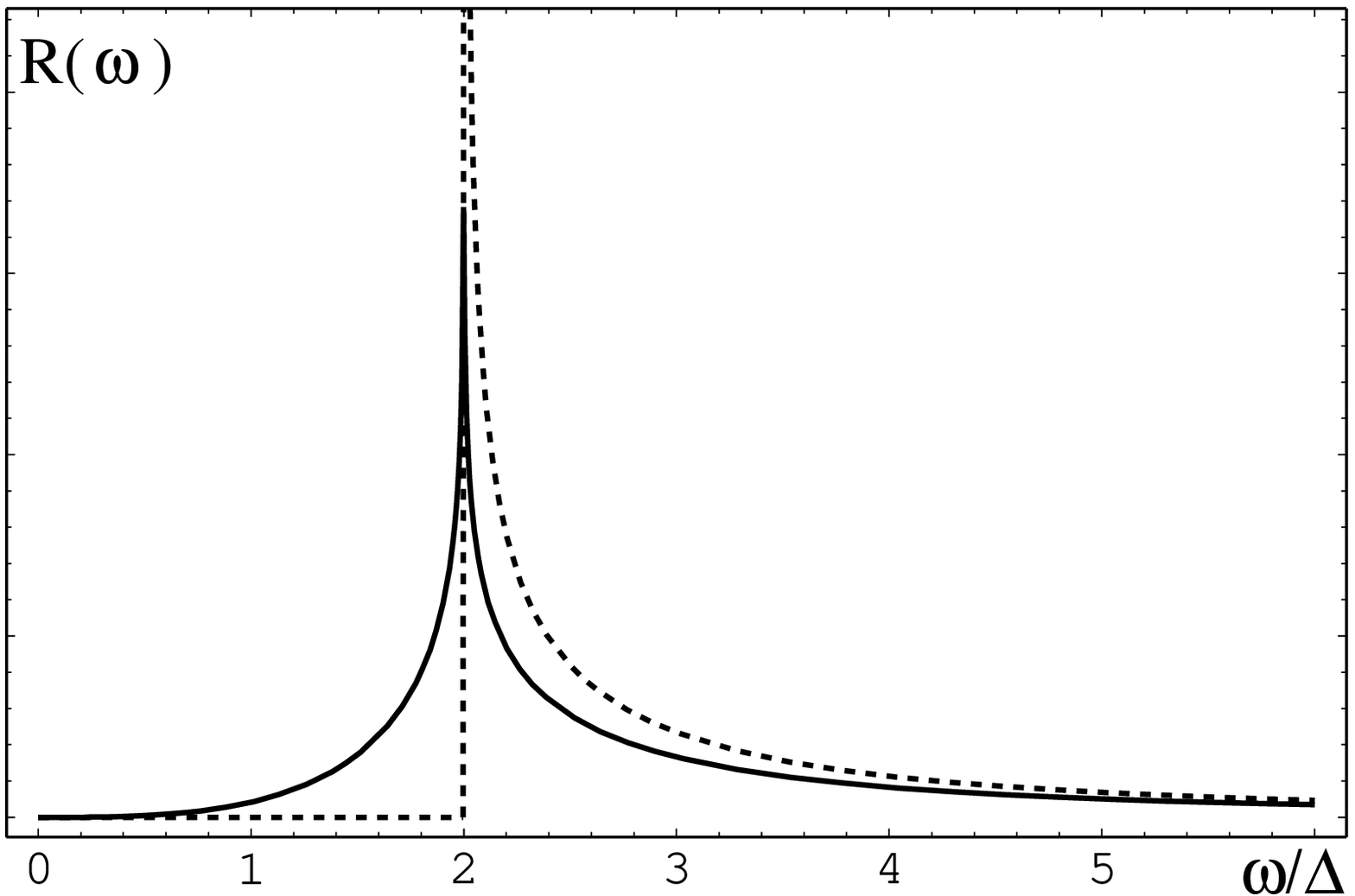}
\end{center}
\caption{The behavior of the Raman response in a BCS superconductor with a
flat gap (dashed line), and for a quadratic variation of the gap near its
maximum (solid line).}
\label{Figure_Raman_gas}
\end{figure}

The behavior of $R(\omega )$ in a gas is shown in Fig.\ref{Figure_Raman_gas}%
. Observe that due to interplay of numerical factors, the logarithmic
singularity shows up only in a near vicinity of $2\Delta $, while at
somewhat larger $\omega $, the angular dependence of the gap becomes
irrelevant, and $R(\omega )$ behaves as $(\omega -2\Delta )^{1/2}$, i.e., as
for a flat gap.

\subsubsection{Raman and optical response at strong coupling}

A nonzero fermionic self-energy mostly affects the optical conductivity for
a simple reason that it becomes finite in the presence of the spin
scattering which can relax fermionic momentum. For a momentum-independent
gap, a finite conductivity emerges above a sharp threshold. This threshold
stems from the fact that at least one of the two fermions in the
conductivity bubble should have a finite $\Sigma ^{\prime \prime }$, i.e.,
its energy should be larger than $\omega _{0}$. Another fermion should be
able to propagate, i.e., its energy should be larger than $\Delta $. The
combination of the two requirements yields the threshold for $\sigma (\omega
>0)$ at $2\Delta +\Delta _{s}$, i.e., at the same frequency where the SIS
tunneling conductance is singular. One can easily demonstrate that for a
flat gap, the conductivity emerges above the threshold as $\epsilon
^{1/2}/\log ^{2}\epsilon $, where, we remind, $\epsilon =\omega -\Delta
-\omega _{0}=\omega -(2\Delta +\Delta _{s})$. This singularity obviously
causes a divergence of the first derivative of the conductivity at $\epsilon
=+0$.

For a true $d-$wave gap, the conductivity is finite for all frequencies
simply because the angular integration in Eq.(\ref{rs}) involves the region
near the nodes, where $\Sigma ^{\prime \prime } $ is nonzero down to the
lowest frequencies. Still, however, we argue that the conductivity is
singular at $\omega _{0}+\Delta $. Indeed, replacing, as before, $\epsilon $
by $\epsilon +\widetilde{\theta }^{2}$, substituting this into the result
for the conductivity for a flat gap, and integrating over $\widetilde{\theta 
}$, we find that for momentum dependent gap, the conductivity itself and its
first derivative are continuous at $\epsilon =0$, but the second derivative
of the conductivity diverges as ${d^2 \sigma}/{d \omega^2} \propto
1/(|\epsilon |\log ^{2}\epsilon)$.

\begin{figure}[tbp]
\begin{center}
\epsfxsize=3.0in \epsfysize=1.8in
\epsffile{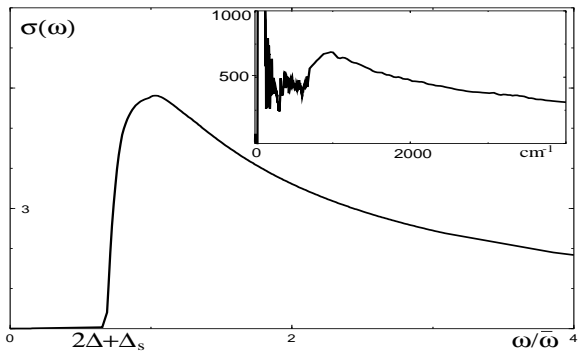}
\end{center}
\caption{Frequency dependence of the optical conductivity computed from the
self energy and pairing vertex determined from the Eliashberg equations
valid near hot spots. The onset of the optical response is $\protect\omega %
=2\Delta +\Delta _{s}$. The contribution from the nodes (not included in
calculations) yields a nonzero conductivity at all $\protect\omega$ and
softens the singularity at $\protect\omega =2\Delta +\Delta _{s}$. The
insert shows the experimental data of Puchkov {\it et al.}~\protect\cite
{basov}. }
\label{Figure_cond_strongc}
\end{figure}

In Fig.\ref{Figure_cond_strongc} we show the result for conductivity
obtained by solving the set of coupled Eliashberg-type equations Eqs.\ref
{setphi}-\ref{setpi}. We clearly see an expected singularity at $2\Delta
+\Delta_s$. Note by passing that at higher frequencies, the theoretical $%
\Sigma(\omega)$ is inversely linear in $\omega$~\cite{rob}.

For the Raman intensity, the strong coupling effects are less relevant.
First, one can prove along the same lines as in previous subsections that
the cubic behavior at low frequencies for $B_{1g}$ scattering (and the
linear behavior for angular independent vertices), and the logarithmic
singularity at $2\Delta $ are general properties of a $d-$wave
superconductor, which survive for all couplings. Thus, similar to the
density of states and the SIS-tunneling spectrum, the Raman response below $%
2\Delta $ is not sensitive to strong coupling effects. Second, near $\omega
_{0}+\Delta $, singular contributions which come from $\Sigma _{+}\Sigma
_{-} $ and $\Phi _{+}\Phi _{-}$ terms in $\Pi _{r}$ in Eq.(\ref{rs}) cancel
each other. As a result, we found that for a flat gap, only the second
derivative of $R(\omega )$ diverges at $\Delta +\omega _{0}$. For a
quadratic variation of a gap near its maximum, the singularity is even
weaker and shows up only in the third derivative of $R(\omega )$. Obviously,
this is a very weak effect, and its determination requires a high quality of
the experiment. Notice, however, that, as we already mentioned, due to the
closeness of hot spots to $(0,\pi )$ and related points, where $v_{F}$
vanishes, the smearing of the singularity due to momentum integration may be
less drastic than in our theory where we used a linearized fermionic
dispersion with some finite Fermi velocity.

\subsubsection{Comparison with experiments}

Our theoretical considerations show that optical measurements are much
better suited to search for the ``fingerprints'' of a magnetic scenario,
then Raman measurements. Evidences for strong coupling effects in the
optical conductivity in superconducting cuprates have been reported in Refs. 
\cite{basov,CSB99,bonn,nuss}. We present the experimental data for $\sigma
(\omega)$ in optimally doped ${\rm YBCO}$ in the inset of Fig.\ref
{Figure_cond_strongc}. We see that the conductivity drops at about $100{\rm %
meV}$. Earlier tunneling measurements of the gap in optimally doped ${\rm %
YBCO}$ yielded $\Delta =29{\rm meV}$~\cite{early}. Combining this with $%
\Delta _{s}=41{\rm meV}$, we find $2\Delta +\Delta _{s}\approx 100{\rm meV}$%
, consistent with the data. We consider this agreement as another argument
in favor of a magnetic scenario.

\begin{figure}[tbp]
\begin{center}
\epsfxsize=2.8in \epsfysize=1.8in
\epsffile{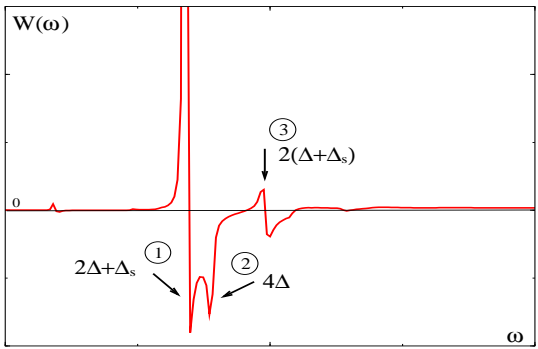}
\end{center}
\caption{ A calculated frequency dependence of $W(\protect\omega )=\frac{%
d^{2}}{d^{2}\protect\omega }(\protect\omega {\rm Re}\protect\sigma ^{-1}(%
\protect\omega ))$. This quantity is a sensitive measure for fine structures
in the optical response.}
\label{Figure_d2sigdw2}
\end{figure}

\begin{figure}[tbp]
\begin{center}
\epsfxsize=3.0in \epsfysize=1.8in
\epsffile{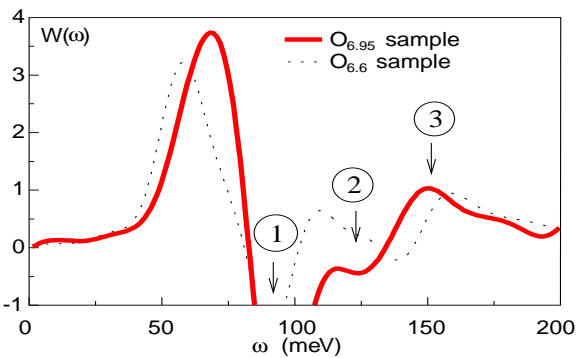}
\end{center}
\caption{ Experimental results for $W(\protect\omega )=\frac{d^{2}}{d^{2}%
\protect\omega }(\protect\omega {\rm Re}\protect\sigma ^{-1}(\protect\omega
))$ from Ref.\protect\cite{CSB99}. The position of the deep minimum agrees
well with $2\Delta +\Delta _{s}$. The extrema at higher frequencies are
consistent with $4\Delta $ and $2\left( \Delta +\Delta _{s}\right) $
predicted by the theory.}
\label{Figure_d2sigdw2_exp}
\end{figure}

\subsubsection{Fine structure of optical conductivity}

We now argue that the measurements of optical conductivity allow one not
only to verify the magnetic scenario, but also to independently determine
both $\Delta _{s}$ and $\Delta $ in the same experiment. We discussed
several times above that in a magnetic scenario, the fermionic self-energy
is singular at two frequencies: at $\omega _{0}=\Delta +\Delta _{s}$, which
is the onset frequency for spin-fluctuation scattering near hot spots, and
at $\omega =3\Delta $, where fermionic damping near hot spots first emerges
due to a direct four-fermion interaction. We argued that in the
spin-fluctuation mechanism, both singularities are due to the same
underlying interaction, and their relative intensity can be obtained within
a model.

In general, the singularity at $3\Delta$ is much weaker at strong coupling,
and can be detected only in the analysis of the derivatives of the fermionic
self-energy. We remind that the singularity in $\Sigma (\omega)$ at $%
\omega_0 $ gives rise to singularity in the conductivity at $\Delta +
\omega_0$, while the $3\Delta$ singularity in $\Sigma (\omega)$ obviously
causes a singularity in conductivity at $\omega = 4\Delta$. Besides, we
should also expect a singularity in $\sigma (\omega)$ at $2\omega_0$, as at
this frequency both fermions in the bubble have a singular $\Sigma
(\omega_0) $.

For phonon superconductors, a fine structure of optical conductivity has
been analyzed by studying a second derivative of conductivity via $W(\omega
)=\frac{d^{2}}{d^{2}\omega }(\omega {\rm Re}\sigma ^{-1}(\omega ))$ which is
proportional to $\alpha^2 (\omega) F(\omega)$ where $\alpha (\omega)$ is an
effective electron-phonon coupling, and $F(\omega)$ is a phonon DOS ~\cite
{a^2f}.

In Fig.\ref{Figure_d2sigdw2} we present the theoretical result for $%
W(\omega) $ in our model \cite{acs2}. First, we clearly see that there is a
sharp maximum in $W(\omega)$ near $2\Delta + \Delta_s$, which is followed by
a deep minimum. This form is consistent with our analytical observation that
for a flat gap (which we used in our numerical analysis), the first
derivative of conductivity diverges at $\omega =2\Delta + \Delta_s$. At a
finite $T$ which is a necessary attribute of a numerical solution, the
singularity is smoothened, and the divergence is transformed into a maximum.
Accordingly, the second derivative of the conductivity should have a maximum
and a minimum near $2\Delta + \Delta_s$. We found from our numerical
analysis that the maximum shifts to lower frequencies with increasing $T$,
but the minimum moves very little from $2\Delta + \Delta_s$, and is
therefore a good measure of a magnetic ``fingerprint''.

Second, we see from Fig.\ref{Figure_d2sigdw2} that besides the maximum and
the minimum near $2\Delta + \Delta_s$, $W(\omega )$ has extra extrema at $%
4\Delta $ and $2\omega _{0} = 2\Delta + 2\Delta_s$. These are precisely the
extra features that we expect, respectively, as a primary effect due to a
singularity in $\Sigma(\omega)$ at $\omega = 3\Delta$, and as a secondary
effect due to a singularity in $\Sigma (\omega)$ at $\omega = \omega_0$.

The experimental result for $W(\omega )$ shown in Fig.\ref
{Figure_d2sigdw2_exp}. We see that the theoretical and experimental plots of 
$W(\omega)$ look rather similar. Furthermore, the relative intensities of
the peaks are at least qualitatively consistent with the theory. Identifying
the extra extrema in the experimental $W(\omega )$ with $4\Delta $ and $%
2\Delta +2\Delta _{s}$, respectively, we obtain $4\Delta \sim 130{\rm meV}$,
and $2\Delta +2\Delta _{s}\sim 150{\rm meV}$. This yields $\Delta \sim 32%
{\rm meV}$, in good agreement with earlier measurements~\cite{Miyakawa99},
and $\Delta _{s}\sim 45{\rm meV}$, which is only slightly larger than the
resonance frequency extracted from neutron measurements~\cite
{neutrons,neutrons2}. Despite the fact that the determination of a second
derivative of a measured quantity is a very subtle procedure, the very
presence of the extra peaks and the fact that their positions fully agree
with the theory, is an indication that the fine structures in the optical
response may indeed be due to strong spin fermion scattering.

\section{comparison with other works}

We now discuss how our work is related to other studies. As we stated in the
very beginning of the paper, the fact that the interaction with a bosonic
mode with frequency $\omega _{0}$ gives rise to a fermionic damping in a
superconductor above $\omega _{0}+\Delta $, is known for conventional $s-$%
wave superconductors.\cite{Scalapino69,carbotte} (see also \cite{varma} and 
\cite{SCM97}). 
Recent angle-integrated photoemission data for lead and niobium~\cite{C2000}
 clearly
demonstrated that the photoemission intensity at low $T$ possesses peak-dip
 features, but the dip is located {\it well above} $3\Delta$. 

The reduction of the spin damping below $2\Delta$ in a $d-$wave
superconductor has been discussed in~\cite{Quinlan94}. It has also been 
argued earlier that the interaction with a nearly resonant collective mode
peaked at ${\bf Q}$ explains the ARPES data. Qualitative arguments for this
have been displayed by Shen and Schrieffer \cite{schr-shen} and Norman and
Ding\cite{Ding}. Norman and Ding also conjectured that the peak-dip
separation may be related to the frequency of a neutron peak, and presented
weak-coupling calculations for a model in which fermions interact with a
resonance bosonic mode. The results of this analysis agree with the ARPES
data.  From this perspective, the novelty of our approach is in that it
presents controlled strong-coupling calculations for a low-energy
spin-fermion model, which verify and extend earlier ideas.

It has been also realized earlier that in a $d-$ wave BCS superconductor,
the dynamical spin response at wave vector ${\bf Q}$ contains an excitonic
pole below $2\Delta $. This has been demonstrated by a number of researchers 
\cite{mazin}. The earlier studies, however, considered a weak coupling
limit, when a bare particle-hole bubble (a building block of RPA series) is
made out of free fermions. However, as we discussed in Sec.~\ref{spo},
performing weak coupling calculations consistently, one finds $\Delta _{s} $
exponentially close to $2\Delta $, and the exponentially small residue of
the resonance peak. Our work extends the idea that the neutron resonance is
a spin exciton to the strong coupling limit.

The results for the density of states, the SIS tunneling conductance, the
Raman intensity, and the optical conductivity in a $d-$wave gas have been
studied in detail by a number of researchers~\cite{maki}. In our approach,
we used these results as a zero-order theory, and considered strong coupling
feedback effects on top of it. As far as we know, for DOS, SIS tunneling and
Raman intensity, these feedback effects have not been studied before. The
form of the optical conductivity and of $W(\omega )$ in cuprates has been
recently studied by Carbotte and co-workers~\cite{CSB99,SCM97,MSC98}. They
also argued that the analysis of $W(\omega )$ supports a magnetic scenario
for the pairing. There is, however, some discrepancy between Ref.~\cite
{CSB99} and out work: it was argued in~\cite{CSB99} that the largest, broad
maximum in $W(\omega )$ is shifted down from $2\Delta +\Delta _{s}$ due to
the angular dependence of the gap, and is located at $\Delta +\Delta _{s}$.
We also found that the broad maximum in $W(\omega )$ is located at a
frequency smaller than $2\Delta +\Delta _{s}$. We however, attribute this
reduction to finite $T$. We showed in the text that for $T\ll \Delta $, the
singularity in $W(\omega )$ in a $d-$wave superconductor is still located
precisely at $2\Delta +\Delta _{s}$.

Finally, several groups ~\cite{sphe} recently suggested phenomenologically
that there is a connection between neutron resonance and specific heat
anomaly in cuprates~\cite{loram}. The verification of this connection within
the spin-fermion model is called for.

\section{Conclusions}

In this paper we demonstrated that the same feedback effect which in the
past allowed one to verify the phononic mechanism of a conventional, $s$%
-wave superconductivity, is also applicable to cuprates, and may allow one
to experimentally detect the ``fingerprints'' of the pairing mechanism in
high $T_c$ superconductors. We argued that for spin-mediated fermionic
scattering, and for the hole-like normal state Fermi surface, the fermionic
spectral function, the density of states, the SIS tunneling conductance, and
the optical conductivity are affected in a certain way by the interaction
with collective spin excitations which in the superconducting state are
propagating, magnon-like quasiparticles with the gap $\Delta _{s}$. We have
shown that the interaction with propagating spin excitations gives rise to
singularities at frequencies $\Delta +\Delta _{s}$ for the spectral function
near hot spots and the DOS, and at $2\Delta +\Delta _{s}$ for the SIS
tunneling conductance and the optical conductivity. We demonstrated that the
value of $\Delta_s$ extracted from these experiments agrees well with the
results of neutron experiments which measure $\Delta_s$ directly.

We further argued that in optical measurements one can also detect
subleading singularities at $4\Delta $ and $2\Delta +2\Delta _{s}$, and that
these fine features have been observed at right frequencies.

We consider the experimental detection of these singularities, particularly
fine structure effects, as the strong evidence in favor of the magnetic
scenario for superconductivity in the cuprates.

It is our pleasure to thank A. M. Finkel'stein and D. Pines for stimulating
discussions on numerous aspects of strong coupling effects in cuprates. We
are also thankful to D. Basov, G. Blumberg, J.C. Campuzano, P. Coleman, L.P.
Gor`kov, P. Johnson, R. Joynt, B. Keimer, D. Khveschenko, G. Kotliar, A.
Millis, M. Norman, S. Sachdev, Q. Si, O. Tchernyshyov, A.Tsvelik, and J.
Zasadzinski for useful conversations. We are also thankful to D. Basov and
J. Zasadzinski for sharing their unpublished results with us. The research
was supported by NSF DMR-9979749 (Ar. A and A. Ch.), and by the Ames
Laboratory, operated for the U.S. Department of Energy by Iowa State
University under contract No. W-7405-Eng-82 (J.S).


\begin{references}
\bibitem{WVH93}  D. A. Wollmann, D. J. Van Harlingen, W. C. Lee, D. M.
Ginsberg, and A. J. Leggett, Phys. Rev. Lett. {\bf 71}, 2134 (1993).

\bibitem{TK94}  C. C. Tsuei, J. R. Kirtley, C. C. Chi, Lock See Yu-Jahnes,
A. Gupta, T. Shaw, J. Z. Sun, and M. B. Ketchen Phys. Rev. Lett. {\bf 73},
593 (1994).

\bibitem{electrond}  
For experimental evidences in favor of $d-$wave pairing in electron-doped cuprates, see
 C.C. Tsuei, J.R. Kirtley, cond-mat/0002341; R.
Prozorov, R. W. Giannetta, P. Fournier, R. L. Greene, cond-mat/0002301.

\bibitem{mahan}  G.D. Mahan, Many-Particle Physics, Plenum Press, 1990.

\bibitem{Scalapino69}  D. J. Scalapino, {\em The electron-phonon interaction
and strong coupling superconductors, }in {\em Superconductivity}, in {\em %
Superconductivity}, Vol. 1, p. 449, Ed. R. D. Parks, Dekker Inc. N.Y. 1969.

\bibitem{pairing}  see e.g., P. Monthoux, A. Balatsky and D. Pines, Phys.
Rev. B {\bf 46}, 14803 (1992); D.J. Scalapino, Phys. Rep. {\bf 250}, 329
(1995).

\bibitem{Norman97}  M. R. Norman {\it et al.}, Phys. Rev. Lett. {\bf 79},
3506 (1997).

\bibitem{shennat}  Z-X. Shen et al, Science {\bf 280}, 259 (1998).

\bibitem{mike}  M. R. Norman, Phys. Rev. B {\bf 61}, 14751 (2000).

\bibitem{commnew}  Strictly speaking, this argumentation is valid only when
the anomalous vertex function $F(\omega)$ is independent on frequency~ \cite
{acsin}. When $F(\omega)$ depends on frequency and possesses the same
singularity as $\Sigma (\omega)$ (as in the solution of the Eliashberg set),
the DOS may even diverge at approaching $\omega_0$ from below. In any
situation, however, it sharply drops above $\omega_0$.

\bibitem{davis}  S. H. Pan, E. W. Hudson, K. M. Lang, H. Eisaki, S. Uchida,
J. C. Davis, Nature, {\bf 403, }746 (2000). More recently, these authors
showed that the tunneling gap has a relatively large spatial variation,
which may be the reason for the broadening of the superconducting peak
observed in photoemission, (J. C. Davis, {\it unpublished}).

\bibitem{Fedorov99}  A.V. Fedorov et al, Phys. Rev. Lett. {\bf 82}, 2179
(1999); T. Valla {\it et al.}, Science, {\bf 285}, 2210 (1999).

\bibitem{Kaminski00}  A. Kaminski et al, Phys. Rev. Lett. {\bf 84}, 1788
(2000).

\bibitem{fisher}  Ch. Renner {\it et al.}, Phys. Rev. Lett. {\bf 80}, 149
(1998); Y. DeWilde et al, ibid {\bf 80}, 153 (1998).

\bibitem{zasad}  J. F. Zasadzinski, L. Ozyuzer, N. Miyakawa, K.E. Gray, D.G.
Hinks and C Kendzora, submitted to Science.

\bibitem{basov}  A. V. Puchkov {\it et al.} J. Phys. Chem. Solids {\bf 59},
1907 (1998); D. N. Basov, R. Liang, B. Dabrovski, D. A. Bonn, W. N. Hardy,
and T. Timusk, Phys. Rev. Lett. {\bf 77}, 4090 (1996).

\bibitem{C2000} A. Chainani {\it et al}, Phys. Rev. Lett. {\bf 85}, 
1966 (2000) and references therein.

\bibitem{CSB99}  J. P. Carbotte, E. Schachinger, D. N. Basov, Nature
(London) {\bf 401}, 354 (1999).

\bibitem{neutrons}  H.F. Fong et al, Phys. Rev. B {\bf 54}, 6708 (1996);

\bibitem{dai}  P. Dai et al, Science {\bf 284}, 1344 (1999).

\bibitem{neutrons2}  H.F. Fong et al, Nature {\bf 398}, 588 (1999).

\bibitem{holstein}  T. Holstein, Phys. Rev. {\bf 96}, 535 (1954); P.B.
Allen, Phys. Rev. B {\bf 3}, 305 (1971).

\bibitem{varma}  P. Littlewood and C.M. Varma, Phys. Rev. B {\bf 46}, 405
(1992).

\bibitem{ssw}  D. J. Scalapino, J. R. Schrieffer and J. W. Wilkins, Phys.
Rev. {\bf 148}, 263 (1966).

\bibitem{Eliashb}  Eliashberg G.M. Sov. Phys. JETP {\bf 11}, 696 (1960).

\bibitem{McMillan69}  W. L. McMillan and J. M. Rowell, {\em Tunneling and
Strong Coupling Superconductivity}, in {\em Superconductivity}, Vol. 1, p.
561, Ed. Parks, Marull, Dekker Inc. N.Y. 1969; W. Shaw and J.C. Swihart,
Phys. Rev. Lett., {\bf 20}, 1000 (1968).

\bibitem{coffey}  D. Coffey, Phys. Rev. B {\bf 42}, 6040 (1990).

\bibitem{migdal}  A.B. Migdal, Sov. Phys. JETP, {\bf 7}, 996 (1958).

\bibitem{Chubukov97}  A. Chubukov, Europhys. Lett. {\bf 44}, 655 (1997).

\bibitem{ac}  Ar. Abanov and A. Chubukov, Phys. Rev. Lett., {\bf 83}, 1652
(1999).

\bibitem{acsin}  Ar. Abanov and A. Chubukov, Phys. Rev. B {\bf 61} ,R9241
(2000).

\bibitem{acf}  Ar. Abanov, A. Chubukov, and A. M. Finkel'stein,
cond-mat/9911445.

\bibitem{acs}  Ar Abanov, A. V. Chubukov, and J. Schmalian, cond-mat/0005163.

\bibitem{oleg}  Ar. Abanov, A. Chubukov, and O. Tchernyshyov, unpublished.

\bibitem{Bourges99}  P. Bourges et al, cond-mat/9902067.

\bibitem{Bourges00}  P. Bourges et al, Science {\bf 288}, 1234 (2000).

\bibitem{Camp}  J. C. Campuzano {\em et al.,} Phys. Rev. Lett {\bf 83}, 3709
(1999).

\bibitem{girsh}  see e.g. A. Chubukov, G. Blumberg, and D. Morr, Solid State
Comm. {\bf 112}, 183 (1999).

\bibitem{schrieffer}  J. R. Schrieffer, {\em Theory of Superconductivity}
(Benjamin, Reading, Mass., 1966).

\bibitem{Devereaux}  T.P. Devereaux {\em et al.}, Phys. Rev. Lett. {\bf 72},
396 (1994); Phys. Rev. B {\bf 54}, 12~523 (1996).

\bibitem{Quinlan96}  S. M. Quinlan, P. J. Hirschfeld, D. J. Scalapino, Phys.
Rev. B {\bf 53}, 8575 (1996).

\bibitem{rob}  R. Haslinger, A. Chubukov, and Ar. Abanov, cond-mat/0009051

\bibitem{a^2f}  B. Farnworth and T. Timusk, Phys. Rev. B {\bf 10}, 
2799 (1974);  F. Marsiglio {\em et al.}, Phys. Lett. A {\bf 245} , 172 (1998).

\bibitem{carbotte}  J. P. Carbotte, Rev. Mod. Phys. {\bf 62}, 1027 (1990).

\bibitem{bonn}  D. A. Bonn, P. Dosanjh, R. Liang, and W. N. Hardy, Phys.
Rev. Lett. {\bf 68}, 2390 (1992).

\bibitem{nuss}  M. C. Nuss, P. M. Mankiewich, M. L. M'OMalley, and E. H.
Westwick, Phys. Rev. Lett. {\bf 66}, 3305 (1991).

\bibitem{early}  I. Maggio-Aprile, Ch. Renner, A. Erb, E. Walker, and \O .
Fischer, Phys. Rev. Lett.{\bf 75}, 2754 (1995).

\bibitem{acs2}  Ar. Abanov, A. Chubukov, and J. Schmalian, in preparation.

\bibitem{Miyakawa99}  N. Miyakawa et al, Phys. Rev. Lett. {\bf 83}, 1018
(1999).

\bibitem{Quinlan94}  B.W. Statt and A. Griffin, Phys. Rev. B {\bf 46}, 3199
(1992); S.M. Quinlan, D.J. Scalapino and N. Bulut, ibid {\bf 49}, 1470
(1994).

\bibitem{schr-shen}  Z-X. Shen and J.R. Schrieffer, Phys. Rev. Lett. {\bf 78}%
, 1771 (1997).

\bibitem{Ding}  M.R. Norman and H. Ding, Phys. Rev. B {\bf 57} R11089
(1998); see also M. Eschrig and M.R. Norman, cond-mat/0005390.

\bibitem{mazin}  D.Z. Liu, Y. Zha and K. Levin, Phys. Rev. Lett. {\bf 75},
4130 (1995); I. Mazin and V. Yakovenko, ibid {\bf 75}, 4134 (1995); C.
Stemmann, C. P\'{e}pin, and M. Lavagna, Phys. Rev. B {\bf 50}, 4075 (1994);
A. Millis and H. Monien, Phys. Rev. B {\bf 54}, 16172 (1996); N. Bulut and
D. Scalapino, Phys. Rev. B {\bf 53}, 5149 (1996);  J. Brinckmann and P.A.
Lee, Phys. Rev. Lett. {\bf 82}, 2915 (1999); S. Sachdev and M. Vojta,
Physica B {\bf 280}, 333 (2000); M.R. Norman, Phys. Rev. B {\bf 61}, 14781
(2000);  O. Tschernyshov, M.R. Norman  and A. Chubukov, cond-mat/0009072; 
D.K. Morr and D. Pines, Phys. Rev. Lett. {\bf 81}, 1086 (1998); F. Onufrieva
and P. Pfeuty, cond-mat/9903097.


\bibitem{maki}  
Ye Sun and K. Maki, Phys. Rev. B {\bf 51}, 6059 (1995) and references
therein.

\bibitem{SCM97}  E. Schachinger, J. P. Carbotte, and F. Marsiglio, Phys.
Rev. B {\bf 56}, 2738 (1997).

\bibitem{MSC98}  F. Marsiglio, T. Startseva, and J. P. Carbotte, Physics
Lett. A {\bf 245}, 172 (1998).

\bibitem{sphe}  D. J. Scalapino and S.R. White, Phys. Rev. B {\bf 58}, 1347
(1998); E. Demler and S.-C. Zhang, Nature, {\ }{\bf 396}, 733 (1999), B.
Janko, cond-mat/9912073.

\bibitem{loram}  J.W. Loram et al, J. Supercond., {\bf 7}, 243 (1994).
\end{references}
\end{document}